\documentclass[journal=jacsat,manuscript=article]{achemso}

\usepackage[T1]{fontenc} 
\usepackage{dcolumn} 

\author{Souren Adhikary$^{1}$, Sasmita Mohakud$^{2}$ and Sudipta Dutta$^{1}$}
\affiliation{$^{1}$Department of Physics, Indian Institute of Science Education and Research (IISER) Tirupati, Tirupati - 517507, Andhra Pradesh, India. \\
 $^{2}$Department of Physics, School of Advanced Sciences, Vellore Institute of Technology, Vellore, Tamil Nadu, 632014, India.}
\email{sdutta@iisertirupati.ac.in}

\title[An \textsf{achemso} demo]
  {Valley-polarization and stable triplet exciton formation in 2D lateral heterostrcuture of hBN-kagome and graphene}

\begin{document}


\date{\today}

\begin{abstract}
Broken spatial inversion symmetry in semiconducting materials with time-reversal pair valleys can exhibit valley polarization. Based on first-principles calculations, here we propose a lateral heterostructure of kagome lattice of hBN and hexagonal graphene domains that exhibits opposite Berry curvature in inequivalent \textbf{K} and \textbf{K}$'$ valleys. Explicit consideration of excitonic scattering processes within $GW$ and Bethe-Salpeter equation formalism confirm insignificant intervalley coupling and consequent valley polarization ability along with 0.46 eV higher binding energy of triplet excitons. Such heterostructure with large charge carrier mobility can be exploited for advanced valleytronic and optoelectronic applications. 

\end{abstract}

\maketitle


The quantum degrees of freedom of electrons can serve as the foundations for information storage and processing\cite{rev-1}. Two-dimensional (2D) hexagonal materials with broken spatial inversion symmetry, possess an additional electronic degree of freedom known as the valley, alongside their charge and spin\cite{rev-2,rev-3}. The primary challenge in harnessing the valley degree of freedom lies in overcoming the valley degeneracy\cite{valley-1,valley-2}. Significant efforts have been dedicated to achieve valley polarization, either through intrinsic mechanisms or by employing extrinsic strategies\cite{Manipulation-1,Manipulation-2,Manipulation-3,Manipulation-4}. For example, the inherent broken inversion symmetry in monolayer transition metal dichalcogenide systems leads to opposite Berry curvature at the \textbf{K} and \textbf{K}$'$ valleys in momentum space\cite{TMD-1,TMD-2,TMD-3}. Consequently, the electrons in these two valleys selectively couple with opposite helicity of circularly polarized light to form excitons\cite{TMD-3}. When a transverse electric field is applied to the system, the movement of charge carriers in the \textbf{K} and \textbf{K}$'$ valleys occurs in opposite directions. These carriers can act as the exciton qubits that is capable of storing information as $`$zero' and $`$one'\cite{valleytronics-1}. Extrinsically, the valley degeneracy can be lifted through applied magnetic fields by few meV\cite{TMD-4,TMD-5,TMD-6}.     

Graphene, the most widely studied 2D system lacks any gap at the valleys and preserves the spatial inversion symmetry that results in zero Berry curvature\cite{neto-rev-modern-physics,geim2007rise}. Creating staggard sublattice potential can serve as a way out to break the inversion symmetry\cite{TMD-1}. In this regard, 2D hexagonal boron nitride (hBN) could act as an ideal candidate for showcasing valley-related properties\cite{hBN-valley-1,hBN-bond-length}. However, due to its weak screening and large insulating gap, the excitons in two valleys couple with each other\cite{hBN-intervalley-prl}. Such 
intervalley coupling significantly supresses the valley-polarization ability of hBN. Therefore, manipulation of spatial inversion symmetry and intervalley scattering in 2D materials are critical steps towards unlocking their potential applications in valley-based information storage.   

Controlled modifications of lattice structures can induce such functionalities in experimentally conducive 2D materials. Recently, 2D kagome lattice comprising of polymerized hetero-triangulene units has been shown computationally to exhibit unique electronic behaviour, featuring a Dirac band enclosed by two flat bands, that can be further engineered to obtain semimetallic or semiconducting properties\cite{kagome-jacs}. These 2D kagome covalent organic frameworks have been successfully synthesized on the Au(111) surface\cite{COF-synthesis-1,COF-synthesis-2}. Such kagome lattice based on graphene has been predicted to show opposite spin localization in adjacent triangulene units and thereby breaking the spatial inversion symmetry\cite{graphene-kagome-1}. A subsequent gap opening at two valleys ensures its valley-polarization ability within first-principles and mean-field level of calculations\cite{graphene-kagome-1}. 

In this letter, we construct the kagome lattice of hBN with subsequent filling of the hexagonal voids by graphene domains to form their lateral heterostructure. This system preserves the time-reversal symmetry with moderate band gap at two valleys and breaks the spatial inversion symmetry. Consequently, this system exhibits opposite Berry curvatures at two valleys with minimal intervalley scattering that are confirmed within density functional theory (DFT) along with $GW$ and Bethe-Salpeter equation (BSE) approximation. The computational details are provided in Supplemental Material (SM)\cite{SM-1}. 
This system also shows higher stabilzation of the triplet excitons over singlet that can be exploited for enhancement of the quantum efficiency for optoelectronic applications. 

\begin{figure}[htb]
\includegraphics[scale=0.4]{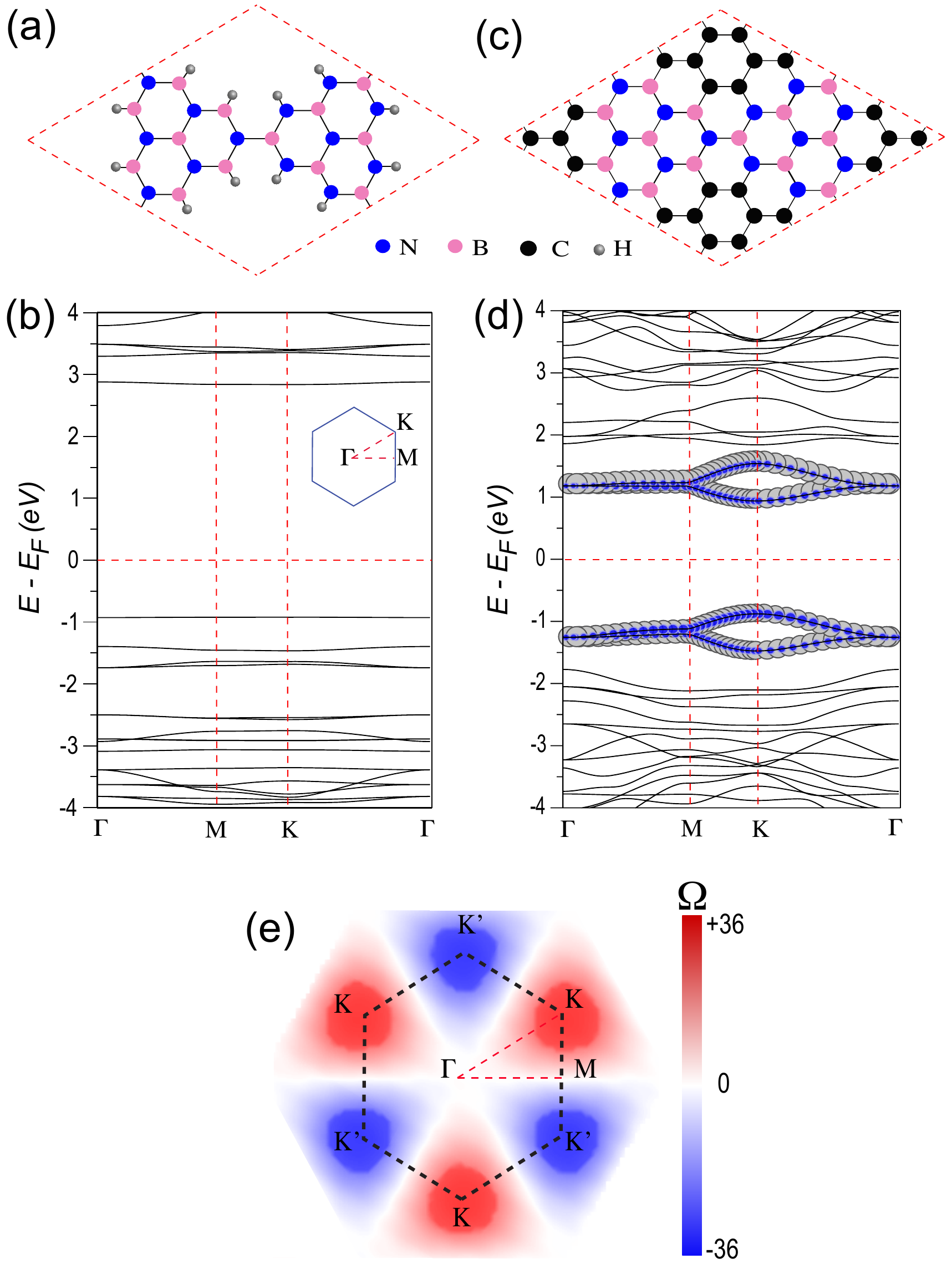}
\caption{\label{fig1} (Color online) (a) The rhombus unit cell of hBNK lattice along with its (b) electronic band structure. The vertical dashed lines indicate high-symmetric points in the hexagonal BZ, shown in the inset. (c) The rhombus unit cell of the lateral heterostructure hBNK-G along with its (d) electronic band structure. The contribution of all carbon atoms and all boron plus nitrogen atoms in the frontier bands are shown as grey and blue (dark) circles, respectively. Both the band structures are obtained within PBE approximation and scaled with respect to the Fermi energy (horizontal dashed lines). (e) The Berry curvature (in atomic units) of the top most valence band of hBNK-G over the hexagonal first BZ (dashed hexagon).}
\end{figure}

First, we construct the kagome lattice hBN, cutting it out from its monolayer and name it hBNK. We passivate the edges by hydrogen atoms to eliminate the dangling bonds to avoid any edge reconstruction. The unit cell of the system in nonmagnetic ground state with lattice constant 12.67 \AA~ is shown in Fig.\ref{fig1}a. The electronic band dispersions of hBNK, as obtained within generalized gradient approximation with Perdew-Burke-Ernzerhof (PBE) exchange and correlation functional are shown in Fig.\ref{fig1}b. The band gap of 4.7 eV in monolayer hBN\cite{hBN-intervalley-prl} significantly reduces to 3.76 eV upon construction of kagome lattice. The bands around the Fermi energy becomes dispersionless, owing to the enhanced electronic localization in presence of hexagonal voids in the lattice. As a result, the charge carrier mobility of hBNK is expected to be very low and may find limited applications in semiconductor devices. Furthermore, these bands can induce enhanced intervalley coupling as has been seen in case of monolayer hBN. The less dispersive bands from $\Gamma$ to \textbf{M} high-symmetry direction in hBN leads to the intervalley coupling and suppresses its valley polarization ability\cite{hBN-intervalley-prl}.

However, such dispersionless bands can be tuned by filling up the hexagonal voids by graphene domains, as shown in Fig.\ref{fig1}c and in Fig.S1. We name this system as hBNK-G that shows nonmagnetic ground state with optimized lattice constant of 12.49 \AA. The minimal lattice mismatch between graphene and hBN allows seamless formation of such lateral heterostructure with precise controll over the domain shape and sizes, as evident from earlier experimental reports\cite{levendorf2012graphene,lattice-mismatch-1,thomas2020step}. Any edge disorder in hBNK can be avoided by formation of such heterostructure and that in turn alters the energy dispersion significantly, as can be seen in Fig.\ref{fig1}d. The hBNK-G system now shows dispersive bands near the Fermi energy with significant reduction of band gap to 1.82 eV at high-symmetry point \textbf{K}, that enables the system for visible spectrum absorption. This substantial change in the energy dispersions can be attributed to the presence of graphene domains that play a dominant role via significant hybridization with the hBNK lattice. This is evident from the projected band structure of hBNK-G in Fig.\ref{fig1}d. The frontier bands show minor contributions from combined B and N atoms and major contributions from C atoms. The enhanced dispersion results in high charge carrier mobility of 1.4×10$^3$ cm$^{2}$V$^{-1}$s$^{-1}$ in hBNK-G\cite{SM-1}, which is comparable with that of black phosphorene\cite{black-phosphorene-carrier-mobility}. Therefore, by creating such lateral heterostructure, one can enhance the optical and semconducting properties. 

\begin{figure*}[htb]
\includegraphics[scale=0.4]{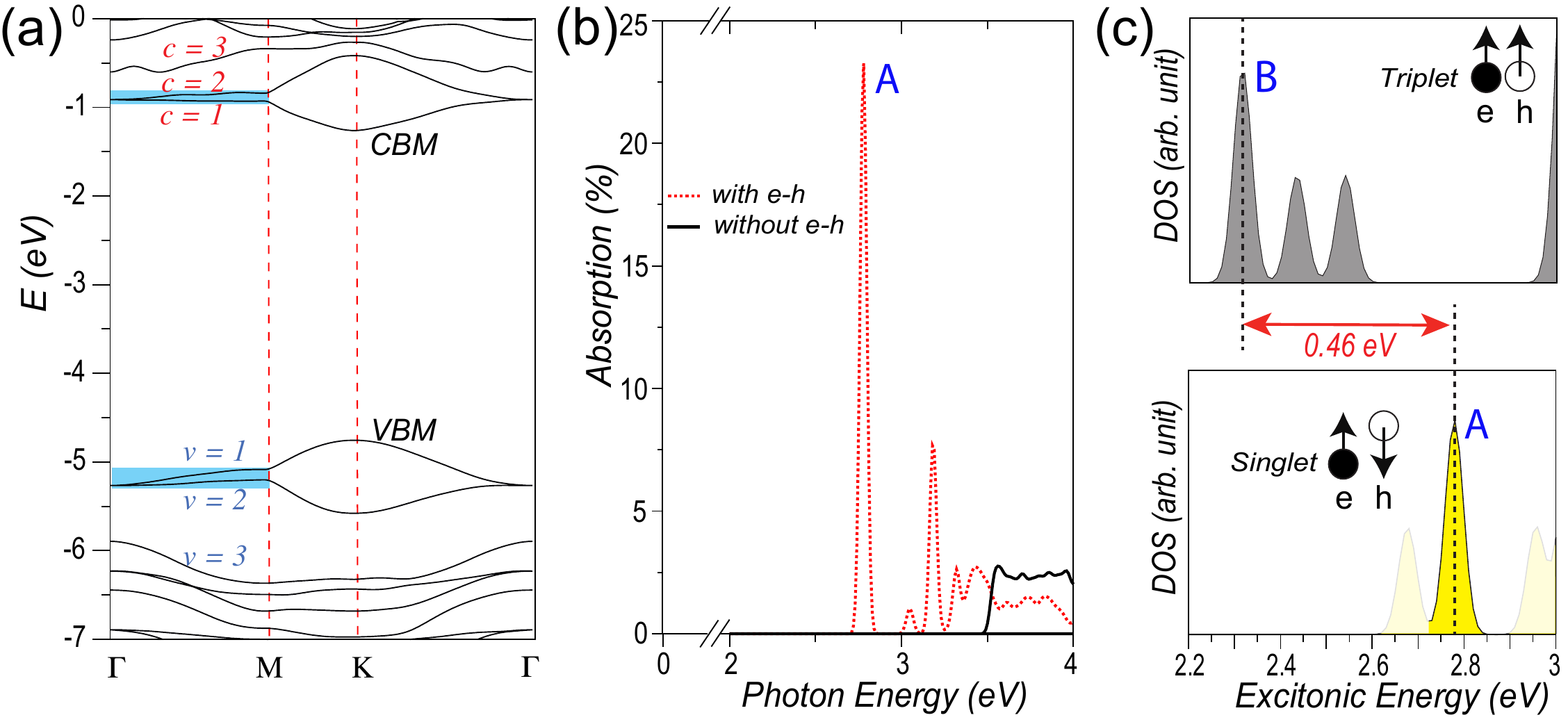}
\caption{\label{fig2} (Color online) (a) The $GW$ band structure of hBNK-G. The valence ($\nu$) and conduction bands ($c$) are marked with numerical values, starting from the frontier bands, denoted by valence band maxima (VBM) and conduction band minima (CBM). The non-dispersive nature of the bands $\nu$ ($c$) = 1 and 2 over $\Gamma$ to \textbf{M} point in first BZ is highlighted with shaded rectangles. (b) The optical absorption spectra of hBNK-G, as obtained with (dotted line) and without (solid line) quasi-electron and quasi-hole interactions. A constant broadening of 20 meV is considered here. The first bright singlet exciton peak is marked with $A$. (c) The DOS of triplet (upper panel) and singlet (lower panel) excitons of the same system. The DOS corresponding to the dark singlets are masked. The gap between the first triplet exciton ($B$) and first bright singlet exciton ($A$) is shown by double-headed arrow.}
\end{figure*}

Furthermore, such enhanced dispersion along with the formation of valleys at time-reversal pair high-symmetric point \textbf{K} and \textbf{K}$'$ in the first BZ can lead to efficient valley polarization in hBNK-G, owing to its broken spatial inversion symmetry, as can be seen in Fig.\ref{fig1}c. This can lead to circular dichroism properties in this system as well. To explore it further, we calculate the Berry curvature for the valence band of hBNK-G using the DFT-based tight binding Hamiltonian and Wannier functions as follows\cite{BC6N-berry}:

\begin{equation}\label{E-1}
\Omega(\textbf{k}) = -\sum_{n\neq n'} 
\frac{2Im <\psi_{n\textbf{k}}|v_x|\psi_{n'\textbf{k}}><\psi_{n'\textbf{k}}|v_y|\psi_{n\textbf{k}}>}{(\varepsilon_n - \varepsilon_{n'})^2}
\end{equation}

\noindent where the summation is over all the occupied valence bands, $v_{x(y)}$ is the velocity operator along the $x$($y$) direction and $\varepsilon_{n}$ is the $n^{th}$ eigenstate energy. From Fig.\ref{fig1}e, it can be seen that the Berry curvature has opposite sign with same magnitude at two valleys, \textbf{K} and \textbf{K}$'$. This leads to an optical selection rule that dictates selective exciations at two valleys under the irradiation of circularly polarized light of different polarities. Moreover, since the Berry curvature behaves analogous to the magnetic field in the momentum space, an inplane electric field generates drift velocities of the similar charge carriers, \textit{i.e.}, electrons or holes in the opposite directions due to a Lorentz-like force. This can be exploited to realize valley-Hall effect in hBNK-G. Note that, the magnitude of the Berry curvature at two valleys comes out to be 36 in Bohr$^2$\cite{SM-1}, which is an order of magnitude higher than that of the superatomic graphene lattice reported earlier\cite{graphene-kagome-1} and comparable to the transition metal dichalcogenide systems\cite{feng2012intrinsic}. 

However, the quasiparticle nature of the excitons and their intervalley scattering can substantially suppress the valley polarization ability, as has been observed in case of 2D hBN system\cite{hBN-intervalley-prl}. Due to reduced Coulomb screening in 2D, such scattering processes play a dominant role and requires a many-body treatment\cite{chernikov2014exciton}.
This has driven us to first investigate the quasiparticle dispersion within $GW$ approximation\cite{GW-method-1} that leads to the following eigenvalue equation, 

\begin{equation}\label{E-2}
[-\frac{1}{2}\nabla^2 + V_{ext} + V_H + \Sigma(\epsilon^{QP}_{n\textbf{k}})] \psi^{QP}_{n\textbf{k}} = \epsilon^{QP}_{n\textbf{k}} \psi^{QP}_{n\textbf{k}}
\end{equation}

\noindent where $\Sigma$ is the self-energy operator within the $GW$ approximation and $\epsilon^{QP}_{n\textbf{k}}$ and $\psi^{QP}_{n\textbf{k}}$ are the quasiparticle energies and wave functions, respectively. The term $V_{ext}$ is potential energy due to the ion-electrons interactions and $V_H$ is the Hartree energy. The electronic self-energy correction results in an enhanced quasiparticle gap of 3.5 eV at the valleys, as can be seen in Fig.\ref{fig2}a. Additional self-consistent update for $G$, \textit{i.e.} $G_1W_0$ results in same gap.
The self-energy correction is highest at $\Gamma$-point with 1.89 eV and lowest at \textbf{K}-point with 1.67 eV, whereas the correction at \textbf{M} point is 1.86 eV.

We further explore the excitonic wave functions within BSE formalism\cite{BSE-1} using the following equation,

\begin{equation}\label{E-3}
(\epsilon^{QP}_{c\textbf{k}} - \epsilon^{QP}_{\nu\textbf{k}})A^S_{\nu c \textbf{k}} +  \sum_{\nu ' c'\textbf{k}'} 
<\nu c \textbf{k}|K^{eh}|\nu ' c'\textbf{k}'> = \Omega^S A^S_{\nu c \textbf{k}}
\end{equation}

\noindent where $\nu$ ($c$) represents the valence (conduction) band index, $A^S_{\nu c \textbf{k}}$ is the envelope function of exciton, $\Omega^S$ excitation energy of exciton state $|S>$, and $K_{eh}$ is the electron-hole interaction kernel. Using the solution of the above equation, the absorption spectra is obtained from the imaginary part of the dielectric function $\varepsilon_2(\omega)$:

\begin{equation}\label{E-4}
\varepsilon_2(\omega) = \frac{16\pi^2e^2}{\omega^2} \sum_S |\textbf{P}.<0|\textbf{v}|S>|^2\delta(\omega -\Omega^S)
\end{equation}

\noindent where $\textbf{P}$ is the polarization of incoming photon, $e$ is the electronic charge, $\textbf{v}$ is the velocity operator, $<0|$ is the fock space within the DFT level, and $|S>$ is the excitonic wave function. The convergence parameters of BSE calculations are provided in the SM. 

We calculate the absorption spectra of hBNK-G in presence of linear polarized light, considering 3 valence and 4 conduction bands to attain the convergence and plot the same in Fig.\ref{fig2}b. The formation of bright singlet exciton peak, A at 2.78 eV indicates a large excitonic binding energy of 0.72 eV with respect to the quasiparticle gap. The excitonic eigenvalue analysis shows that this peak arises from two degenerate excitonic states (see Fig.S4). This degeneracy can be broken by probing circularly polarized light of defined chirality. We further plot the density of states (DOS) corresponding to the singlet and triplet excitons in Fig.\ref{fig2}c, which reveals a large singlet-triplet gap of 0.46 eV with more stable triplet exciton. Note that, below the DOS corresponding to the bright singlet exciton peak $A$, there appears a dark singlet exciton, characterized by negligible oscillator strength as compared to the bright one. That is why, the singlet-triplet splitting is calculated with respect to the bright singlet exciton \textit{i.e}, peak $A$. 

\begin{figure}[h] 
\includegraphics[scale=0.4]{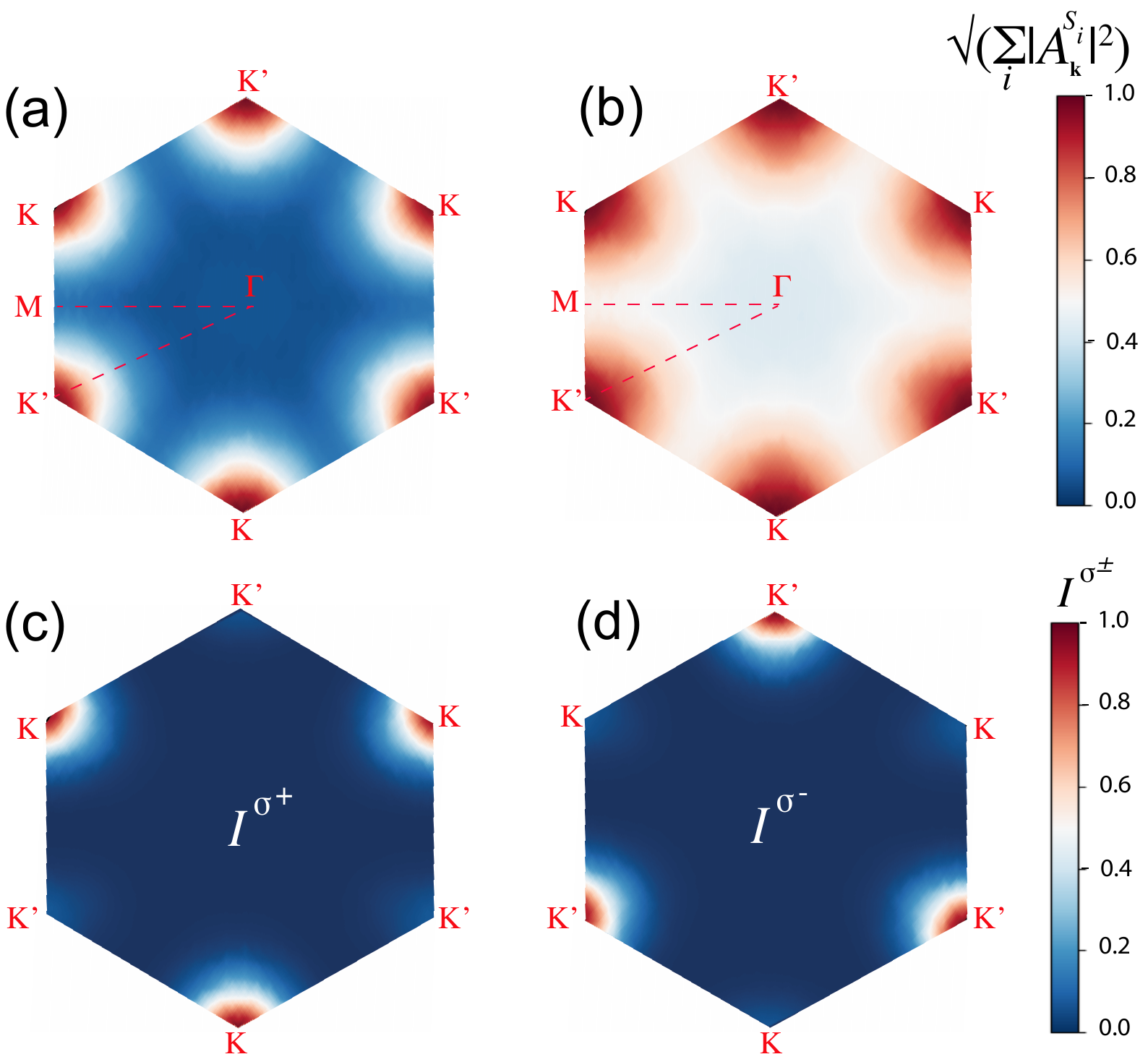}
\caption{\label{fig3} (Color online) The square root of the sum of the squared excitonic envelope
function $A^{S_i}_{\textbf{k}}$ of (a) the first bright singlet ($A$) and (b) the first triplet ($B$) exciton of hBNK-G over the hexagonal first BZ. The color bar displays the normalized value of the same. The oscillator strengths of the first bright singlet ($A$) exciton in presence of (c) left- and (d) right-handed circular polarized lights. The color bar depicts the normalized values of the oscillator strength.}
\end{figure}

To investigate this large splitting, we explore the optical transitions among few valence and conduction bands that are marked with numerical indices in Fig.\ref{fig2}a. We calculate the total excitation amplitude in between two band indices using the equation: $W_{tot} = \sum_{\textbf{k}}|A_{\nu c\textbf{k}}|^2$, where $A_{\nu c\textbf{k}}$ is the amplitude of the particular excitation. We present the corresponding data in Table.I. We find that the singlet excitons get formed majorly due to the transition from the top of the valence band ($\nu$ = 1) to the bottom of the conduction band ($c$ = 1). However, excitations among bands, $\nu$ = 1, 2 and $c$ = 1, 2 show substantial contributions towards the triplet exciton. To gain further insight, we investigate the \textbf{k}-resolved optical transitions for both the first bright singlet (peak A) and first triplet (peak B) excitons in terms of the \textbf{k}-resolved envelope function and plot $\sqrt{\sum_i|A^{S_i}_{\textbf{k}}|^2}$ in Fig.\ref{fig3}a \& b, respectively. Note that, here $i$ is the degeneracy of the excitonic states. As can be seen, the singlet exciton is getting formed due to the optical transitions at and near the \textbf{K} and \textbf{K}$'$ valleys (see Fig.\ref{fig3}a). Significant contributions to the triplet excitons are arising from these valleys too (see Fig.\ref{fig3}b). This can be attributed to the attractive direct electron-hole screened Coulomb interaction term in the interaction kernel, which contributes towards the formation of both singlet and triplet excitons. Further investigation reveals that, the triplet exciton formation at valleys is a consequence of only $\nu$ = 1 to $c$ = 1 transitions (see Fig.S7b). In addition to this, the optical transitions among the other bands towards the triplet formation happen majorly over $\Gamma$ - \textbf{M} path (see Fig.S7-c), where these bands are mostly dispersionless, as highlighted in Fig.\ref{fig2}a. Such flat band induced triplet exciton formation has been reported in superatomic graphene lattice\cite{sethi2021flat}.

\begin{table}[h]
\caption{\label{tab:table1}
The contributions (in \%) from band-to-band optical transitions towards the formation of the first bright singlet ($A$) and first triplet ($B$) excitonic peaks. The values are rounded-off upto the first oder.}
\begin{tabular}{cccccccc}
Singlet &$c$ =1 &$c$ =2&$c$ =3&
Triplet &$c$ =1 &$c$ =2&$c$ =3\\ 
\hline
$\nu$ =1& 97.9 & 0.5 & 0.1 &$\nu$ =1
& 77.3 & 03.4 & 0.0 \\
$\nu$ =1& 00.4 & 0.4 & 0.1 &$\nu$ =2
& 03.7 & 15.6&0.0 \\
$\nu$ =1& 00.2 & 0.1 & 0.1 &$\nu$ =3
& 00.0 & 00.0 & 0.0 \\
\end{tabular}
\end{table}
   
The envelope function plots in Fig.\ref{fig3}a \& b further indicate the extent of intervalley coupling for singlet and triplet excitons, respectively. The triplet excitons experience strong intervalley scattering, as evident from higher magnitude of envelope function in between two valleys. However, in case of the bright singlet exciton, the intervalley coupling between \textbf{K} and \textbf{K}$'$ valleys comes out to be only 12.5\%. Higher magnitude of opposite Berry curvature at these two valleys along with such minimal intervalley coupling indicates efficient valley polarization ability of hBNK-G system. For further confirmation, we calculate the oscillator strengths of the bright singlet excitation under the irradiation of circularly polarized lights of different chirality over the first BZ, using the following expression\cite{hBN-intervalley-prl,xiao2012coupled,mahon2019quantum},

\begin{equation}\label{E-6}
I^{\sigma^\pm}\sim \sum_i|A^{S_i}_{\textbf{k}}\textbf{P}.<\nu\textbf{k}|(v_x\pm iv_y)\hat{a}|c\textbf{k}>|^2
\end{equation}

\noindent where $\sigma^\pm$ indicate different chiralities of the circular polarized light, $v_x$ ($v_y$) is the component of the velocity operator in \textit{x}(\textit{y})-direction and $\hat{a}$ is the corresponding unit vector. As mentioned before, the doubly degenerate eignestates corresponding to the bright singlet exciton get splitted while interacting with the left ($\sigma^+$) and right ($\sigma^-$) circular polarized lights and we plot the corresponding oscillator strengths in Fig.\ref{fig3}c \& d, respectively. As can be seen, the $\sigma^+$ forms excitons selectively at \textbf{K} valley. In presence of in-plane electric field, the electrons and holes in this valley will prefer to move in opposite directions, owing to opposite Berry curvatures of the valence and conduction bands. However, the combination of $\sigma^-$ light with in-plane electric field will drive the charge carriers at \textbf{K}$'$ valley in the opposite directions as compared to the other valley, owing to the sign reversal of the Berry curvature (see Fig.\ref{fig1}e). Therefore, the hBNK-G system can exhibit efficient circular dichroism valley-Hall effect.

To summarize, based on first-principle calculations, we propose the 2D lateral heterostructure of kagome lattice of hBN and hexagonal graphene domains that can exhibit promising electronic and optical properties. An optical gap of 2.78 eV with 0.72 eV binding energy for singlet excitons and high charge carrier mobility of 1.4 $\times$ 10$^3$ cm$^{2}$V$^{-1}$s$^{-1}$ indicates efficient optoelectronic properties of this hBNK-G system. Furthermore, this system is capable of stabilizing the triplet exciton over the singlet one by 0.46 eV that can enhance exciton life-time and the quantum efficiency of optical absorption\cite{triplet-stability-1,triplet-stability-2}. Owing to its broken spatial inversion symmetry, this system exhibits opposite Berry curvature at time-reversal pair valleys, \textbf{K} and \textbf{K}$'$ with negligible intervalley scattering that is confirmed within many-body BSE calculations. Consequently, the charge carriers at each valley can be excited selectively by controlling the chirality of the circular polarized lights, as shown explicitly in terms of their oscillator strengths. This property can be exploited for realizing the excitonic qubits through circular-dichroism valley-Hall device. Our study shows new pathways of inducing interesting optical properties by constructing kagome-based periodic lateral heterostructure of widely used 2D materials like graphene and hBN.

SA and SD thank IISER Tirupati for Intramural Funding and Science and Engineering Research Board, Dept. of Science and Technology, Govt. of India for research grant (CRG/2021/001731). The authors thank Prof. K. Wakabayashi (Kwansei Gakuin University) for illuminating discussions. The authors acknowledge National Supercomputing Mission (NSM) for providing computing resources of ‘PARAM Brahma’ at IISER Pune, which is implemented by C-DAC and supported by the Ministry of Electronics and Information Technology (MeitY) and DST, Govt. of India.

\nocite{*}

\clearpage

\begin{thebibliography}{40}%
\makeatletter
\providecommand \@ifxundefined [1]{%
 \@ifx{#1\undefined}
}%
\providecommand \@ifnum [1]{%
 \ifnum #1\expandafter \@firstoftwo
 \else \expandafter \@secondoftwo
 \fi
}%
\providecommand \@ifx [1]{%
 \ifx #1\expandafter \@firstoftwo
 \else \expandafter \@secondoftwo
 \fi
}%
\providecommand \natexlab [1]{#1}%
\providecommand \enquote  [1]{``#1''}%
\providecommand \bibnamefont  [1]{#1}%
\providecommand \bibfnamefont [1]{#1}%
\providecommand \citenamefont [1]{#1}%
\providecommand \href@noop [0]{\@secondoftwo}%
\providecommand \href [0]{\begingroup \@sanitize@url \@href}%
\providecommand \@href[1]{\@@startlink{#1}\@@href}%
\providecommand \@@href[1]{\endgroup#1\@@endlink}%
\providecommand \@sanitize@url [0]{\catcode `\\12\catcode `\$12\catcode
  `\&12\catcode `\#12\catcode `\^12\catcode `\_12\catcode `\%12\relax}%
\providecommand \@@startlink[1]{}%
\providecommand \@@endlink[0]{}%
\providecommand \url  [0]{\begingroup\@sanitize@url \@url }%
\providecommand \@url [1]{\endgroup\@href {#1}{\urlprefix }}%
\providecommand \urlprefix  [0]{URL }%
\providecommand \Eprint [0]{\href }%
\providecommand \doibase [0]{https://doi.org/}%
\providecommand \selectlanguage [0]{\@gobble}%
\providecommand \bibinfo  [0]{\@secondoftwo}%
\providecommand \bibfield  [0]{\@secondoftwo}%
\providecommand \translation [1]{[#1]}%
\providecommand \BibitemOpen [0]{}%
\providecommand \bibitemStop [0]{}%
\providecommand \bibitemNoStop [0]{.\EOS\space}%
\providecommand \EOS [0]{\spacefactor3000\relax}%
\providecommand \BibitemShut  [1]{\csname bibitem#1\endcsname}%
\let\auto@bib@innerbib\@empty
\bibitem [{\citenamefont {Mak}\ \emph {et~al.}(2018)\citenamefont {Mak},
  \citenamefont {Xiao},\ and\ \citenamefont {Shan}}]{rev-1}%
  \BibitemOpen
  \bibfield  {author} {\bibinfo {author} {\bibfnamefont {K.~F.}\ \bibnamefont
  {Mak}}, \bibinfo {author} {\bibfnamefont {D.}~\bibnamefont {Xiao}},\ and\
  \bibinfo {author} {\bibfnamefont {J.}~\bibnamefont {Shan}},\ }\href@noop {}
  {\bibfield  {journal} {\bibinfo  {journal} {Nat. Photonics}\ }\textbf
  {\bibinfo {volume} {12}},\ \bibinfo {pages} {451} (\bibinfo {year}
  {2018})}\BibitemShut {NoStop}%
\bibitem [{\citenamefont {Mak}\ \emph {et~al.}(2014)\citenamefont {Mak},
  \citenamefont {McGill}, \citenamefont {Park},\ and\ \citenamefont
  {McEuen}}]{rev-2}%
  \BibitemOpen
  \bibfield  {author} {\bibinfo {author} {\bibfnamefont {K.~F.}\ \bibnamefont
  {Mak}}, \bibinfo {author} {\bibfnamefont {K.~L.}\ \bibnamefont {McGill}},
  \bibinfo {author} {\bibfnamefont {J.}~\bibnamefont {Park}},\ and\ \bibinfo
  {author} {\bibfnamefont {P.~L.}\ \bibnamefont {McEuen}},\ }\href@noop {}
  {\bibfield  {journal} {\bibinfo  {journal} {Science}\ }\textbf {\bibinfo
  {volume} {344}},\ \bibinfo {pages} {1489} (\bibinfo {year}
  {2014})}\BibitemShut {NoStop}%
\bibitem [{\citenamefont {Wu}\ \emph {et~al.}(2013)\citenamefont {Wu},
  \citenamefont {Ross}, \citenamefont {Liu}, \citenamefont {Aivazian},
  \citenamefont {Jones}, \citenamefont {Fei}, \citenamefont {Zhu},
  \citenamefont {Xiao}, \citenamefont {Yao}, \citenamefont {Cobden} \emph
  {et~al.}}]{rev-3}%
  \BibitemOpen
  \bibfield  {author} {\bibinfo {author} {\bibfnamefont {S.}~\bibnamefont
  {Wu}}, \bibinfo {author} {\bibfnamefont {J.~S.}\ \bibnamefont {Ross}},
  \bibinfo {author} {\bibfnamefont {G.-B.}\ \bibnamefont {Liu}}, \bibinfo
  {author} {\bibfnamefont {G.}~\bibnamefont {Aivazian}}, \bibinfo {author}
  {\bibfnamefont {A.}~\bibnamefont {Jones}}, \bibinfo {author} {\bibfnamefont
  {Z.}~\bibnamefont {Fei}}, \bibinfo {author} {\bibfnamefont {W.}~\bibnamefont
  {Zhu}}, \bibinfo {author} {\bibfnamefont {D.}~\bibnamefont {Xiao}}, \bibinfo
  {author} {\bibfnamefont {W.}~\bibnamefont {Yao}}, \bibinfo {author}
  {\bibfnamefont {D.}~\bibnamefont {Cobden}}, \emph {et~al.},\ }\href@noop {}
  {\bibfield  {journal} {\bibinfo  {journal} {Nat. Phys.}\ }\textbf {\bibinfo
  {volume} {9}},\ \bibinfo {pages} {149} (\bibinfo {year} {2013})}\BibitemShut
  {NoStop}%
\bibitem [{\citenamefont {Ye}\ \emph {et~al.}(2017)\citenamefont {Ye},
  \citenamefont {Sun},\ and\ \citenamefont {Heinz}}]{valley-1}%
  \BibitemOpen
  \bibfield  {author} {\bibinfo {author} {\bibfnamefont {Z.}~\bibnamefont
  {Ye}}, \bibinfo {author} {\bibfnamefont {D.}~\bibnamefont {Sun}},\ and\
  \bibinfo {author} {\bibfnamefont {T.~F.}\ \bibnamefont {Heinz}},\ }\href@noop
  {} {\bibfield  {journal} {\bibinfo  {journal} {Nat. Phys.}\ }\textbf
  {\bibinfo {volume} {13}},\ \bibinfo {pages} {26} (\bibinfo {year}
  {2017})}\BibitemShut {NoStop}%
\bibitem [{\citenamefont {Zeng}\ \emph {et~al.}(2012)\citenamefont {Zeng},
  \citenamefont {Dai}, \citenamefont {Yao}, \citenamefont {Xiao},\ and\
  \citenamefont {Cui}}]{valley-2}%
  \BibitemOpen
  \bibfield  {author} {\bibinfo {author} {\bibfnamefont {H.}~\bibnamefont
  {Zeng}}, \bibinfo {author} {\bibfnamefont {J.}~\bibnamefont {Dai}}, \bibinfo
  {author} {\bibfnamefont {W.}~\bibnamefont {Yao}}, \bibinfo {author}
  {\bibfnamefont {D.}~\bibnamefont {Xiao}},\ and\ \bibinfo {author}
  {\bibfnamefont {X.}~\bibnamefont {Cui}},\ }\href@noop {} {\bibfield
  {journal} {\bibinfo  {journal} {Nat. Nanotechnol.}\ }\textbf {\bibinfo
  {volume} {7}},\ \bibinfo {pages} {490} (\bibinfo {year} {2012})}\BibitemShut
  {NoStop}%
\bibitem [{\citenamefont {Aivazian}\ \emph
  {et~al.}(2015{\natexlab{a}})\citenamefont {Aivazian}, \citenamefont {Gong},
  \citenamefont {Jones}, \citenamefont {Chu}, \citenamefont {Yan},
  \citenamefont {Mandrus}, \citenamefont {Zhang}, \citenamefont {Cobden},
  \citenamefont {Yao},\ and\ \citenamefont {Xu}}]{Manipulation-1}%
  \BibitemOpen
  \bibfield  {author} {\bibinfo {author} {\bibfnamefont {G.}~\bibnamefont
  {Aivazian}}, \bibinfo {author} {\bibfnamefont {Z.}~\bibnamefont {Gong}},
  \bibinfo {author} {\bibfnamefont {A.~M.}\ \bibnamefont {Jones}}, \bibinfo
  {author} {\bibfnamefont {R.-L.}\ \bibnamefont {Chu}}, \bibinfo {author}
  {\bibfnamefont {J.}~\bibnamefont {Yan}}, \bibinfo {author} {\bibfnamefont
  {D.~G.}\ \bibnamefont {Mandrus}}, \bibinfo {author} {\bibfnamefont
  {C.}~\bibnamefont {Zhang}}, \bibinfo {author} {\bibfnamefont
  {D.}~\bibnamefont {Cobden}}, \bibinfo {author} {\bibfnamefont
  {W.}~\bibnamefont {Yao}},\ and\ \bibinfo {author} {\bibfnamefont
  {X.}~\bibnamefont {Xu}},\ }\href@noop {} {\bibfield  {journal} {\bibinfo
  {journal} {Nat. Phys.}\ }\textbf {\bibinfo {volume} {11}},\ \bibinfo {pages}
  {148} (\bibinfo {year} {2015}{\natexlab{a}})}\BibitemShut {NoStop}%
\bibitem [{\citenamefont {MacNeill}\ \emph
  {et~al.}(2015{\natexlab{a}})\citenamefont {MacNeill}, \citenamefont {Heikes},
  \citenamefont {Mak}, \citenamefont {Anderson}, \citenamefont {Korm{\'a}nyos},
  \citenamefont {Z{\'o}lyomi}, \citenamefont {Park},\ and\ \citenamefont
  {Ralph}}]{Manipulation-2}%
  \BibitemOpen
  \bibfield  {author} {\bibinfo {author} {\bibfnamefont {D.}~\bibnamefont
  {MacNeill}}, \bibinfo {author} {\bibfnamefont {C.}~\bibnamefont {Heikes}},
  \bibinfo {author} {\bibfnamefont {K.~F.}\ \bibnamefont {Mak}}, \bibinfo
  {author} {\bibfnamefont {Z.}~\bibnamefont {Anderson}}, \bibinfo {author}
  {\bibfnamefont {A.}~\bibnamefont {Korm{\'a}nyos}}, \bibinfo {author}
  {\bibfnamefont {V.}~\bibnamefont {Z{\'o}lyomi}}, \bibinfo {author}
  {\bibfnamefont {J.}~\bibnamefont {Park}},\ and\ \bibinfo {author}
  {\bibfnamefont {D.~C.}\ \bibnamefont {Ralph}},\ }\href@noop {} {\bibfield
  {journal} {\bibinfo  {journal} {Phys. Rev. Lett.}\ }\textbf {\bibinfo
  {volume} {114}},\ \bibinfo {pages} {037401} (\bibinfo {year}
  {2015}{\natexlab{a}})}\BibitemShut {NoStop}%
\bibitem [{\citenamefont {Xu}\ \emph {et~al.}(2018)\citenamefont {Xu},
  \citenamefont {Yang}, \citenamefont {Shen}, \citenamefont {Zhou},
  \citenamefont {Zhu},\ and\ \citenamefont {Feng}}]{Manipulation-3}%
  \BibitemOpen
  \bibfield  {author} {\bibinfo {author} {\bibfnamefont {L.}~\bibnamefont
  {Xu}}, \bibinfo {author} {\bibfnamefont {M.}~\bibnamefont {Yang}}, \bibinfo
  {author} {\bibfnamefont {L.}~\bibnamefont {Shen}}, \bibinfo {author}
  {\bibfnamefont {J.}~\bibnamefont {Zhou}}, \bibinfo {author} {\bibfnamefont
  {T.}~\bibnamefont {Zhu}},\ and\ \bibinfo {author} {\bibfnamefont {Y.~P.}\
  \bibnamefont {Feng}},\ }\href@noop {} {\bibfield  {journal} {\bibinfo
  {journal} {Phys. Rev. B}\ }\textbf {\bibinfo {volume} {97}},\ \bibinfo
  {pages} {041405(R)} (\bibinfo {year} {2018})}\BibitemShut {NoStop}%
\bibitem [{\citenamefont {Hu}\ \emph {et~al.}(2020)\citenamefont {Hu},
  \citenamefont {Zhao}, \citenamefont {Gao}, \citenamefont {Wu}, \citenamefont
  {Hong}, \citenamefont {Stroppa},\ and\ \citenamefont {Ren}}]{Manipulation-4}%
  \BibitemOpen
  \bibfield  {author} {\bibinfo {author} {\bibfnamefont {T.}~\bibnamefont
  {Hu}}, \bibinfo {author} {\bibfnamefont {G.}~\bibnamefont {Zhao}}, \bibinfo
  {author} {\bibfnamefont {H.}~\bibnamefont {Gao}}, \bibinfo {author}
  {\bibfnamefont {Y.}~\bibnamefont {Wu}}, \bibinfo {author} {\bibfnamefont
  {J.}~\bibnamefont {Hong}}, \bibinfo {author} {\bibfnamefont {A.}~\bibnamefont
  {Stroppa}},\ and\ \bibinfo {author} {\bibfnamefont {W.}~\bibnamefont {Ren}},\
  }\href@noop {} {\bibfield  {journal} {\bibinfo  {journal} {Phys. Rev. B}\
  }\textbf {\bibinfo {volume} {101}},\ \bibinfo {pages} {125401} (\bibinfo
  {year} {2020})}\BibitemShut {NoStop}%
\bibitem [{\citenamefont {Xiao}\ \emph {et~al.}(2007)\citenamefont {Xiao},
  \citenamefont {Yao},\ and\ \citenamefont {Niu}}]{TMD-1}%
  \BibitemOpen
  \bibfield  {author} {\bibinfo {author} {\bibfnamefont {D.}~\bibnamefont
  {Xiao}}, \bibinfo {author} {\bibfnamefont {W.}~\bibnamefont {Yao}},\ and\
  \bibinfo {author} {\bibfnamefont {Q.}~\bibnamefont {Niu}},\ }\href@noop {}
  {\bibfield  {journal} {\bibinfo  {journal} {Phys. Rev. Lett.}\ }\textbf
  {\bibinfo {volume} {99}},\ \bibinfo {pages} {236809} (\bibinfo {year}
  {2007})}\BibitemShut {NoStop}%
\bibitem [{\citenamefont {Xiao}\ \emph
  {et~al.}(2012{\natexlab{a}})\citenamefont {Xiao}, \citenamefont {Liu},
  \citenamefont {Feng}, \citenamefont {Xu},\ and\ \citenamefont {Yao}}]{TMD-2}%
  \BibitemOpen
  \bibfield  {author} {\bibinfo {author} {\bibfnamefont {D.}~\bibnamefont
  {Xiao}}, \bibinfo {author} {\bibfnamefont {G.-B.}\ \bibnamefont {Liu}},
  \bibinfo {author} {\bibfnamefont {W.}~\bibnamefont {Feng}}, \bibinfo {author}
  {\bibfnamefont {X.}~\bibnamefont {Xu}},\ and\ \bibinfo {author}
  {\bibfnamefont {W.}~\bibnamefont {Yao}},\ }\href@noop {} {\bibfield
  {journal} {\bibinfo  {journal} {Phys. Rev. Lett.}\ }\textbf {\bibinfo
  {volume} {108}},\ \bibinfo {pages} {196802} (\bibinfo {year}
  {2012}{\natexlab{a}})}\BibitemShut {NoStop}%
\bibitem [{\citenamefont {Cao}\ \emph {et~al.}(2012)\citenamefont {Cao},
  \citenamefont {Wang}, \citenamefont {Han}, \citenamefont {Ye}, \citenamefont
  {Zhu}, \citenamefont {Shi}, \citenamefont {Niu}, \citenamefont {Tan},
  \citenamefont {Wang}, \citenamefont {Liu} \emph {et~al.}}]{TMD-3}%
  \BibitemOpen
  \bibfield  {author} {\bibinfo {author} {\bibfnamefont {T.}~\bibnamefont
  {Cao}}, \bibinfo {author} {\bibfnamefont {G.}~\bibnamefont {Wang}}, \bibinfo
  {author} {\bibfnamefont {W.}~\bibnamefont {Han}}, \bibinfo {author}
  {\bibfnamefont {H.}~\bibnamefont {Ye}}, \bibinfo {author} {\bibfnamefont
  {C.}~\bibnamefont {Zhu}}, \bibinfo {author} {\bibfnamefont {J.}~\bibnamefont
  {Shi}}, \bibinfo {author} {\bibfnamefont {Q.}~\bibnamefont {Niu}}, \bibinfo
  {author} {\bibfnamefont {P.}~\bibnamefont {Tan}}, \bibinfo {author}
  {\bibfnamefont {E.}~\bibnamefont {Wang}}, \bibinfo {author} {\bibfnamefont
  {B.}~\bibnamefont {Liu}}, \emph {et~al.},\ }\href@noop {} {\bibfield
  {journal} {\bibinfo  {journal} {Nat. Commun.}\ }\textbf {\bibinfo {volume}
  {3}},\ \bibinfo {pages} {887} (\bibinfo {year} {2012})}\BibitemShut {NoStop}%
\bibitem [{\citenamefont {Pacchioni}(2020)}]{valleytronics-1}%
  \BibitemOpen
  \bibfield  {author} {\bibinfo {author} {\bibfnamefont {G.}~\bibnamefont
  {Pacchioni}},\ }\href@noop {} {\bibfield  {journal} {\bibinfo  {journal}
  {Nat. Rev. Mater.}\ }\textbf {\bibinfo {volume} {5}},\ \bibinfo {pages} {480}
  (\bibinfo {year} {2020})}\BibitemShut {NoStop}%
\bibitem [{\citenamefont {Srivastava}\ \emph {et~al.}(2015)\citenamefont
  {Srivastava}, \citenamefont {Sidler}, \citenamefont {Allain}, \citenamefont
  {Lembke}, \citenamefont {Kis},\ and\ \citenamefont {Imamo{\u{g}}lu}}]{TMD-4}%
  \BibitemOpen
  \bibfield  {author} {\bibinfo {author} {\bibfnamefont {A.}~\bibnamefont
  {Srivastava}}, \bibinfo {author} {\bibfnamefont {M.}~\bibnamefont {Sidler}},
  \bibinfo {author} {\bibfnamefont {A.~V.}\ \bibnamefont {Allain}}, \bibinfo
  {author} {\bibfnamefont {D.~S.}\ \bibnamefont {Lembke}}, \bibinfo {author}
  {\bibfnamefont {A.}~\bibnamefont {Kis}},\ and\ \bibinfo {author}
  {\bibfnamefont {A.}~\bibnamefont {Imamo{\u{g}}lu}},\ }\href@noop {}
  {\bibfield  {journal} {\bibinfo  {journal} {Nat. Phys.}\ }\textbf {\bibinfo
  {volume} {11}},\ \bibinfo {pages} {141} (\bibinfo {year} {2015})}\BibitemShut
  {NoStop}%
\bibitem [{\citenamefont {Aivazian}\ \emph
  {et~al.}(2015{\natexlab{b}})\citenamefont {Aivazian}, \citenamefont {Gong},
  \citenamefont {Jones}, \citenamefont {Chu}, \citenamefont {Yan},
  \citenamefont {Mandrus}, \citenamefont {Zhang}, \citenamefont {Cobden},
  \citenamefont {Yao},\ and\ \citenamefont {Xu}}]{TMD-5}%
  \BibitemOpen
  \bibfield  {author} {\bibinfo {author} {\bibfnamefont {G.}~\bibnamefont
  {Aivazian}}, \bibinfo {author} {\bibfnamefont {Z.}~\bibnamefont {Gong}},
  \bibinfo {author} {\bibfnamefont {A.~M.}\ \bibnamefont {Jones}}, \bibinfo
  {author} {\bibfnamefont {R.-L.}\ \bibnamefont {Chu}}, \bibinfo {author}
  {\bibfnamefont {J.}~\bibnamefont {Yan}}, \bibinfo {author} {\bibfnamefont
  {D.~G.}\ \bibnamefont {Mandrus}}, \bibinfo {author} {\bibfnamefont
  {C.}~\bibnamefont {Zhang}}, \bibinfo {author} {\bibfnamefont
  {D.}~\bibnamefont {Cobden}}, \bibinfo {author} {\bibfnamefont
  {W.}~\bibnamefont {Yao}},\ and\ \bibinfo {author} {\bibfnamefont
  {X.}~\bibnamefont {Xu}},\ }\href@noop {} {\bibfield  {journal} {\bibinfo
  {journal} {Nat. Phys.}\ }\textbf {\bibinfo {volume} {11}},\ \bibinfo {pages}
  {148} (\bibinfo {year} {2015}{\natexlab{b}})}\BibitemShut {NoStop}%
\bibitem [{\citenamefont {MacNeill}\ \emph
  {et~al.}(2015{\natexlab{b}})\citenamefont {MacNeill}, \citenamefont {Heikes},
  \citenamefont {Mak}, \citenamefont {Anderson}, \citenamefont {Korm{\'a}nyos},
  \citenamefont {Z{\'o}lyomi}, \citenamefont {Park},\ and\ \citenamefont
  {Ralph}}]{TMD-6}%
  \BibitemOpen
  \bibfield  {author} {\bibinfo {author} {\bibfnamefont {D.}~\bibnamefont
  {MacNeill}}, \bibinfo {author} {\bibfnamefont {C.}~\bibnamefont {Heikes}},
  \bibinfo {author} {\bibfnamefont {K.~F.}\ \bibnamefont {Mak}}, \bibinfo
  {author} {\bibfnamefont {Z.}~\bibnamefont {Anderson}}, \bibinfo {author}
  {\bibfnamefont {A.}~\bibnamefont {Korm{\'a}nyos}}, \bibinfo {author}
  {\bibfnamefont {V.}~\bibnamefont {Z{\'o}lyomi}}, \bibinfo {author}
  {\bibfnamefont {J.}~\bibnamefont {Park}},\ and\ \bibinfo {author}
  {\bibfnamefont {D.~C.}\ \bibnamefont {Ralph}},\ }\href@noop {} {\bibfield
  {journal} {\bibinfo  {journal} {Phys. Rev. Lett.}\ }\textbf {\bibinfo
  {volume} {114}},\ \bibinfo {pages} {037401} (\bibinfo {year}
  {2015}{\natexlab{b}})}\BibitemShut {NoStop}%
\bibitem [{\citenamefont {Neto}\ \emph {et~al.}(2009)\citenamefont {Neto},
  \citenamefont {Guinea}, \citenamefont {Peres}, \citenamefont {Novoselov},\
  and\ \citenamefont {Geim}}]{neto-rev-modern-physics}%
  \BibitemOpen
  \bibfield  {author} {\bibinfo {author} {\bibfnamefont {A.~C.}\ \bibnamefont
  {Neto}}, \bibinfo {author} {\bibfnamefont {F.}~\bibnamefont {Guinea}},
  \bibinfo {author} {\bibfnamefont {N.~M.}\ \bibnamefont {Peres}}, \bibinfo
  {author} {\bibfnamefont {K.~S.}\ \bibnamefont {Novoselov}},\ and\ \bibinfo
  {author} {\bibfnamefont {A.~K.}\ \bibnamefont {Geim}},\ }\href@noop {}
  {\bibfield  {journal} {\bibinfo  {journal} {Rev . Mod. Phys.}\ }\textbf
  {\bibinfo {volume} {81}},\ \bibinfo {pages} {109} (\bibinfo {year}
  {2009})}\BibitemShut {NoStop}%
\bibitem [{\citenamefont {Geim}\ and\ \citenamefont
  {Novoselov}(2007)}]{geim2007rise}%
  \BibitemOpen
  \bibfield  {author} {\bibinfo {author} {\bibfnamefont {A.~K.}\ \bibnamefont
  {Geim}}\ and\ \bibinfo {author} {\bibfnamefont {K.~S.}\ \bibnamefont
  {Novoselov}},\ }\href@noop {} {\bibfield  {journal} {\bibinfo  {journal}
  {Nat. Mater.}\ }\textbf {\bibinfo {volume} {6}},\ \bibinfo {pages} {183}
  (\bibinfo {year} {2007})}\BibitemShut {NoStop}%
\bibitem [{\citenamefont {Song}\ \emph {et~al.}(2017)\citenamefont {Song},
  \citenamefont {Li}, \citenamefont {Wang}, \citenamefont {Bai}, \citenamefont
  {Wang}, \citenamefont {Du}, \citenamefont {Liu}, \citenamefont {Wang},
  \citenamefont {Han}, \citenamefont {Yang} \emph {et~al.}}]{hBN-valley-1}%
  \BibitemOpen
  \bibfield  {author} {\bibinfo {author} {\bibfnamefont {Z.}~\bibnamefont
  {Song}}, \bibinfo {author} {\bibfnamefont {Z.}~\bibnamefont {Li}}, \bibinfo
  {author} {\bibfnamefont {H.}~\bibnamefont {Wang}}, \bibinfo {author}
  {\bibfnamefont {X.}~\bibnamefont {Bai}}, \bibinfo {author} {\bibfnamefont
  {W.}~\bibnamefont {Wang}}, \bibinfo {author} {\bibfnamefont {H.}~\bibnamefont
  {Du}}, \bibinfo {author} {\bibfnamefont {S.}~\bibnamefont {Liu}}, \bibinfo
  {author} {\bibfnamefont {C.}~\bibnamefont {Wang}}, \bibinfo {author}
  {\bibfnamefont {J.}~\bibnamefont {Han}}, \bibinfo {author} {\bibfnamefont
  {Y.}~\bibnamefont {Yang}}, \emph {et~al.},\ }\href@noop {} {\bibfield
  {journal} {\bibinfo  {journal} {Nano Lett.}\ }\textbf {\bibinfo {volume}
  {17}},\ \bibinfo {pages} {2079} (\bibinfo {year} {2017})}\BibitemShut
  {NoStop}%
\bibitem [{\citenamefont {Elias}\ \emph {et~al.}(2019)\citenamefont {Elias},
  \citenamefont {Valvin}, \citenamefont {Pelini}, \citenamefont {Summerfield},
  \citenamefont {Mellor}, \citenamefont {Cheng}, \citenamefont {Eaves},
  \citenamefont {Foxon}, \citenamefont {Beton}, \citenamefont {Novikov} \emph
  {et~al.}}]{hBN-bond-length}%
  \BibitemOpen
  \bibfield  {author} {\bibinfo {author} {\bibfnamefont {C.}~\bibnamefont
  {Elias}}, \bibinfo {author} {\bibfnamefont {P.}~\bibnamefont {Valvin}},
  \bibinfo {author} {\bibfnamefont {T.}~\bibnamefont {Pelini}}, \bibinfo
  {author} {\bibfnamefont {A.}~\bibnamefont {Summerfield}}, \bibinfo {author}
  {\bibfnamefont {C.}~\bibnamefont {Mellor}}, \bibinfo {author} {\bibfnamefont
  {T.}~\bibnamefont {Cheng}}, \bibinfo {author} {\bibfnamefont
  {L.}~\bibnamefont {Eaves}}, \bibinfo {author} {\bibfnamefont
  {C.}~\bibnamefont {Foxon}}, \bibinfo {author} {\bibfnamefont
  {P.}~\bibnamefont {Beton}}, \bibinfo {author} {\bibfnamefont
  {S.}~\bibnamefont {Novikov}}, \emph {et~al.},\ }\href@noop {} {\bibfield
  {journal} {\bibinfo  {journal} {Nat. Commun.}\ }\textbf {\bibinfo {volume}
  {10}},\ \bibinfo {pages} {2639} (\bibinfo {year} {2019})}\BibitemShut
  {NoStop}%
\bibitem [{\citenamefont {Zhang}\ \emph {et~al.}(2022)\citenamefont {Zhang},
  \citenamefont {Ong}, \citenamefont {Ruan}, \citenamefont {Wu}, \citenamefont
  {Shi}, \citenamefont {Tang},\ and\ \citenamefont
  {Louie}}]{hBN-intervalley-prl}%
  \BibitemOpen
  \bibfield  {author} {\bibinfo {author} {\bibfnamefont {F.}~\bibnamefont
  {Zhang}}, \bibinfo {author} {\bibfnamefont {C.~S.}\ \bibnamefont {Ong}},
  \bibinfo {author} {\bibfnamefont {J.~W.}\ \bibnamefont {Ruan}}, \bibinfo
  {author} {\bibfnamefont {M.}~\bibnamefont {Wu}}, \bibinfo {author}
  {\bibfnamefont {X.~Q.}\ \bibnamefont {Shi}}, \bibinfo {author} {\bibfnamefont
  {Z.~K.}\ \bibnamefont {Tang}},\ and\ \bibinfo {author} {\bibfnamefont
  {S.~G.}\ \bibnamefont {Louie}},\ }\href@noop {} {\bibfield  {journal}
  {\bibinfo  {journal} {Phys. Rev. Lett.}\ }\textbf {\bibinfo {volume} {128}},\
  \bibinfo {pages} {047402} (\bibinfo {year} {2022})}\BibitemShut {NoStop}%
\bibitem [{\citenamefont {Jing}\ and\ \citenamefont
  {Heine}(2018)}]{kagome-jacs}%
  \BibitemOpen
  \bibfield  {author} {\bibinfo {author} {\bibfnamefont {Y.}~\bibnamefont
  {Jing}}\ and\ \bibinfo {author} {\bibfnamefont {T.}~\bibnamefont {Heine}},\
  }\href@noop {} {\bibfield  {journal} {\bibinfo  {journal} {J. Am. Chem.
  Soc.}\ }\textbf {\bibinfo {volume} {141}},\ \bibinfo {pages} {743} (\bibinfo
  {year} {2018})}\BibitemShut {NoStop}%
\bibitem [{\citenamefont {Pavli{\v{c}}ek}\ \emph {et~al.}(2017)\citenamefont
  {Pavli{\v{c}}ek}, \citenamefont {Mistry}, \citenamefont {Majzik},
  \citenamefont {Moll}, \citenamefont {Meyer}, \citenamefont {Fox},\ and\
  \citenamefont {Gross}}]{COF-synthesis-1}%
  \BibitemOpen
  \bibfield  {author} {\bibinfo {author} {\bibfnamefont {N.}~\bibnamefont
  {Pavli{\v{c}}ek}}, \bibinfo {author} {\bibfnamefont {A.}~\bibnamefont
  {Mistry}}, \bibinfo {author} {\bibfnamefont {Z.}~\bibnamefont {Majzik}},
  \bibinfo {author} {\bibfnamefont {N.}~\bibnamefont {Moll}}, \bibinfo {author}
  {\bibfnamefont {G.}~\bibnamefont {Meyer}}, \bibinfo {author} {\bibfnamefont
  {D.~J.}\ \bibnamefont {Fox}},\ and\ \bibinfo {author} {\bibfnamefont
  {L.}~\bibnamefont {Gross}},\ }\href@noop {} {\bibfield  {journal} {\bibinfo
  {journal} {Nat. Nanotechno.}\ }\textbf {\bibinfo {volume} {12}},\ \bibinfo
  {pages} {308} (\bibinfo {year} {2017})}\BibitemShut {NoStop}%
\bibitem [{\citenamefont {Galeotti}\ \emph {et~al.}(2020)\citenamefont
  {Galeotti}, \citenamefont {De~Marchi}, \citenamefont {Hamzehpoor},
  \citenamefont {MacLean}, \citenamefont {Rajeswara~Rao}, \citenamefont {Chen},
  \citenamefont {Besteiro}, \citenamefont {Dettmann}, \citenamefont {Ferrari},
  \citenamefont {Frezza} \emph {et~al.}}]{COF-synthesis-2}%
  \BibitemOpen
  \bibfield  {author} {\bibinfo {author} {\bibfnamefont {G.}~\bibnamefont
  {Galeotti}}, \bibinfo {author} {\bibfnamefont {F.}~\bibnamefont {De~Marchi}},
  \bibinfo {author} {\bibfnamefont {E.}~\bibnamefont {Hamzehpoor}}, \bibinfo
  {author} {\bibfnamefont {O.}~\bibnamefont {MacLean}}, \bibinfo {author}
  {\bibfnamefont {M.}~\bibnamefont {Rajeswara~Rao}}, \bibinfo {author}
  {\bibfnamefont {Y.}~\bibnamefont {Chen}}, \bibinfo {author} {\bibfnamefont
  {L.}~\bibnamefont {Besteiro}}, \bibinfo {author} {\bibfnamefont
  {D.}~\bibnamefont {Dettmann}}, \bibinfo {author} {\bibfnamefont
  {L.}~\bibnamefont {Ferrari}}, \bibinfo {author} {\bibfnamefont
  {F.}~\bibnamefont {Frezza}}, \emph {et~al.},\ }\href@noop {} {\bibfield
  {journal} {\bibinfo  {journal} {Nat. Mater.}\ }\textbf {\bibinfo {volume}
  {19}},\ \bibinfo {pages} {874} (\bibinfo {year} {2020})}\BibitemShut
  {NoStop}%
\bibitem [{\citenamefont {Zhou}\ and\ \citenamefont
  {Liu}(2020)}]{graphene-kagome-1}%
  \BibitemOpen
  \bibfield  {author} {\bibinfo {author} {\bibfnamefont {Y.}~\bibnamefont
  {Zhou}}\ and\ \bibinfo {author} {\bibfnamefont {F.}~\bibnamefont {Liu}},\
  }\href@noop {} {\bibfield  {journal} {\bibinfo  {journal} {Nano Lett.}\
  }\textbf {\bibinfo {volume} {21}},\ \bibinfo {pages} {230} (\bibinfo {year}
  {2020})}\BibitemShut {NoStop}%
\bibitem [{SM-()}]{SM-1}%
  \BibitemOpen
  \href@noop {} {\bibinfo  {journal} {See Supplemental Material for
  computational details and parameterization along with more details of triplet
  exciton formation}\ }\BibitemShut {NoStop}%
\bibitem [{\citenamefont {Levendorf}\ \emph {et~al.}(2012)\citenamefont
  {Levendorf}, \citenamefont {Kim}, \citenamefont {Brown}, \citenamefont
  {Huang}, \citenamefont {Havener}, \citenamefont {Muller},\ and\ \citenamefont
  {Park}}]{levendorf2012graphene}%
  \BibitemOpen
\bibfield  {journal} {  }\bibfield  {author} {\bibinfo {author} {\bibfnamefont
  {M.~P.}\ \bibnamefont {Levendorf}}, \bibinfo {author} {\bibfnamefont {C.-J.}\
  \bibnamefont {Kim}}, \bibinfo {author} {\bibfnamefont {L.}~\bibnamefont
  {Brown}}, \bibinfo {author} {\bibfnamefont {P.~Y.}\ \bibnamefont {Huang}},
  \bibinfo {author} {\bibfnamefont {R.~W.}\ \bibnamefont {Havener}}, \bibinfo
  {author} {\bibfnamefont {D.~A.}\ \bibnamefont {Muller}},\ and\ \bibinfo
  {author} {\bibfnamefont {J.}~\bibnamefont {Park}},\ }\href@noop {} {\bibfield
   {journal} {\bibinfo  {journal} {Nature}\ }\textbf {\bibinfo {volume}
  {488}},\ \bibinfo {pages} {627} (\bibinfo {year} {2012})}\BibitemShut
  {NoStop}%
\bibitem [{\citenamefont {Liu}\ \emph {et~al.}(2013)\citenamefont {Liu},
  \citenamefont {Ma}, \citenamefont {Shi}, \citenamefont {Zhou}, \citenamefont
  {Gong}, \citenamefont {Lei}, \citenamefont {Yang}, \citenamefont {Zhang},
  \citenamefont {Yu}, \citenamefont {Hackenberg} \emph
  {et~al.}}]{lattice-mismatch-1}%
  \BibitemOpen
  \bibfield  {author} {\bibinfo {author} {\bibfnamefont {Z.}~\bibnamefont
  {Liu}}, \bibinfo {author} {\bibfnamefont {L.}~\bibnamefont {Ma}}, \bibinfo
  {author} {\bibfnamefont {G.}~\bibnamefont {Shi}}, \bibinfo {author}
  {\bibfnamefont {W.}~\bibnamefont {Zhou}}, \bibinfo {author} {\bibfnamefont
  {Y.}~\bibnamefont {Gong}}, \bibinfo {author} {\bibfnamefont {S.}~\bibnamefont
  {Lei}}, \bibinfo {author} {\bibfnamefont {X.}~\bibnamefont {Yang}}, \bibinfo
  {author} {\bibfnamefont {J.}~\bibnamefont {Zhang}}, \bibinfo {author}
  {\bibfnamefont {J.}~\bibnamefont {Yu}}, \bibinfo {author} {\bibfnamefont
  {K.~P.}\ \bibnamefont {Hackenberg}}, \emph {et~al.},\ }\href@noop {}
  {\bibfield  {journal} {\bibinfo  {journal} {Nat. Nanotechno.}\ }\textbf
  {\bibinfo {volume} {8}},\ \bibinfo {pages} {119} (\bibinfo {year}
  {2013})}\BibitemShut {NoStop}%
\bibitem [{\citenamefont {Thomas}\ \emph {et~al.}(2020)\citenamefont {Thomas},
  \citenamefont {Bradford}, \citenamefont {Cheng}, \citenamefont {Summerfield},
  \citenamefont {Wrigley}, \citenamefont {Mellor}, \citenamefont {Khlobystov},
  \citenamefont {Foxon}, \citenamefont {Eaves}, \citenamefont {Novikov} \emph
  {et~al.}}]{thomas2020step}%
  \BibitemOpen
  \bibfield  {author} {\bibinfo {author} {\bibfnamefont {J.}~\bibnamefont
  {Thomas}}, \bibinfo {author} {\bibfnamefont {J.}~\bibnamefont {Bradford}},
  \bibinfo {author} {\bibfnamefont {T.~S.}\ \bibnamefont {Cheng}}, \bibinfo
  {author} {\bibfnamefont {A.}~\bibnamefont {Summerfield}}, \bibinfo {author}
  {\bibfnamefont {J.}~\bibnamefont {Wrigley}}, \bibinfo {author} {\bibfnamefont
  {C.~J.}\ \bibnamefont {Mellor}}, \bibinfo {author} {\bibfnamefont {A.~N.}\
  \bibnamefont {Khlobystov}}, \bibinfo {author} {\bibfnamefont {C.~T.}\
  \bibnamefont {Foxon}}, \bibinfo {author} {\bibfnamefont {L.}~\bibnamefont
  {Eaves}}, \bibinfo {author} {\bibfnamefont {S.~V.}\ \bibnamefont {Novikov}},
  \emph {et~al.},\ }\href@noop {} {\bibfield  {journal} {\bibinfo  {journal}
  {2D Mater.}\ }\textbf {\bibinfo {volume} {7}},\ \bibinfo {pages} {035014}
  (\bibinfo {year} {2020})}\BibitemShut {NoStop}%
\bibitem [{\citenamefont {Qiao}\ \emph {et~al.}(2014)\citenamefont {Qiao},
  \citenamefont {Kong}, \citenamefont {Hu}, \citenamefont {Yang},\ and\
  \citenamefont {Ji}}]{black-phosphorene-carrier-mobility}%
  \BibitemOpen
  \bibfield  {author} {\bibinfo {author} {\bibfnamefont {J.}~\bibnamefont
  {Qiao}}, \bibinfo {author} {\bibfnamefont {X.}~\bibnamefont {Kong}}, \bibinfo
  {author} {\bibfnamefont {Z.-X.}\ \bibnamefont {Hu}}, \bibinfo {author}
  {\bibfnamefont {F.}~\bibnamefont {Yang}},\ and\ \bibinfo {author}
  {\bibfnamefont {W.}~\bibnamefont {Ji}},\ }\href@noop {} {\bibfield  {journal}
  {\bibinfo  {journal} {Nat. Commun.}\ }\textbf {\bibinfo {volume} {5}},\
  \bibinfo {pages} {4475} (\bibinfo {year} {2014})}\BibitemShut {NoStop}%
\bibitem [{\citenamefont {Liu}\ \emph {et~al.}(2018)\citenamefont {Liu},
  \citenamefont {Ma}, \citenamefont {Gao}, \citenamefont {Zhang}, \citenamefont
  {Ai}, \citenamefont {Li},\ and\ \citenamefont {Zhao}}]{BC6N-berry}%
  \BibitemOpen
  \bibfield  {author} {\bibinfo {author} {\bibfnamefont {X.}~\bibnamefont
  {Liu}}, \bibinfo {author} {\bibfnamefont {X.}~\bibnamefont {Ma}}, \bibinfo
  {author} {\bibfnamefont {H.}~\bibnamefont {Gao}}, \bibinfo {author}
  {\bibfnamefont {X.}~\bibnamefont {Zhang}}, \bibinfo {author} {\bibfnamefont
  {H.}~\bibnamefont {Ai}}, \bibinfo {author} {\bibfnamefont {W.}~\bibnamefont
  {Li}},\ and\ \bibinfo {author} {\bibfnamefont {M.}~\bibnamefont {Zhao}},\
  }\href@noop {} {\bibfield  {journal} {\bibinfo  {journal} {Nanoscale}\
  }\textbf {\bibinfo {volume} {10}},\ \bibinfo {pages} {13179} (\bibinfo {year}
  {2018})}\BibitemShut {NoStop}%
\bibitem [{\citenamefont {Feng}\ \emph {et~al.}(2012)\citenamefont {Feng},
  \citenamefont {Yao}, \citenamefont {Zhu}, \citenamefont {Zhou}, \citenamefont
  {Yao},\ and\ \citenamefont {Xiao}}]{feng2012intrinsic}%
  \BibitemOpen
  \bibfield  {author} {\bibinfo {author} {\bibfnamefont {W.}~\bibnamefont
  {Feng}}, \bibinfo {author} {\bibfnamefont {Y.}~\bibnamefont {Yao}}, \bibinfo
  {author} {\bibfnamefont {W.}~\bibnamefont {Zhu}}, \bibinfo {author}
  {\bibfnamefont {J.}~\bibnamefont {Zhou}}, \bibinfo {author} {\bibfnamefont
  {W.}~\bibnamefont {Yao}},\ and\ \bibinfo {author} {\bibfnamefont
  {D.}~\bibnamefont {Xiao}},\ }\href@noop {} {\bibfield  {journal} {\bibinfo
  {journal} {Phys. Rev. B}\ }\textbf {\bibinfo {volume} {86}},\ \bibinfo
  {pages} {165108} (\bibinfo {year} {2012})}\BibitemShut {NoStop}%
\bibitem [{\citenamefont {Chernikov}\ \emph {et~al.}(2014)\citenamefont
  {Chernikov}, \citenamefont {Berkelbach}, \citenamefont {Hill}, \citenamefont
  {Rigosi}, \citenamefont {Li}, \citenamefont {Aslan}, \citenamefont
  {Reichman}, \citenamefont {Hybertsen},\ and\ \citenamefont
  {Heinz}}]{chernikov2014exciton}%
  \BibitemOpen
  \bibfield  {author} {\bibinfo {author} {\bibfnamefont {A.}~\bibnamefont
  {Chernikov}}, \bibinfo {author} {\bibfnamefont {T.~C.}\ \bibnamefont
  {Berkelbach}}, \bibinfo {author} {\bibfnamefont {H.~M.}\ \bibnamefont
  {Hill}}, \bibinfo {author} {\bibfnamefont {A.}~\bibnamefont {Rigosi}},
  \bibinfo {author} {\bibfnamefont {Y.}~\bibnamefont {Li}}, \bibinfo {author}
  {\bibfnamefont {B.}~\bibnamefont {Aslan}}, \bibinfo {author} {\bibfnamefont
  {D.~R.}\ \bibnamefont {Reichman}}, \bibinfo {author} {\bibfnamefont {M.~S.}\
  \bibnamefont {Hybertsen}},\ and\ \bibinfo {author} {\bibfnamefont {T.~F.}\
  \bibnamefont {Heinz}},\ }\href@noop {} {\bibfield  {journal} {\bibinfo
  {journal} {Phys. Rev. Lett.}\ }\textbf {\bibinfo {volume} {113}},\ \bibinfo
  {pages} {076802} (\bibinfo {year} {2014})}\BibitemShut {NoStop}%
\bibitem [{\citenamefont {Hybertsen}\ and\ \citenamefont
  {Louie}(1986)}]{GW-method-1}%
  \BibitemOpen
  \bibfield  {author} {\bibinfo {author} {\bibfnamefont {M.~S.}\ \bibnamefont
  {Hybertsen}}\ and\ \bibinfo {author} {\bibfnamefont {S.~G.}\ \bibnamefont
  {Louie}},\ }\href@noop {} {\bibfield  {journal} {\bibinfo  {journal} {Phys.
  Rev. B}\ }\textbf {\bibinfo {volume} {34}},\ \bibinfo {pages} {5390}
  (\bibinfo {year} {1986})}\BibitemShut {NoStop}%
\bibitem [{\citenamefont {Rohlfing}\ and\ \citenamefont {Louie}(2000)}]{BSE-1}%
  \BibitemOpen
  \bibfield  {author} {\bibinfo {author} {\bibfnamefont {M.}~\bibnamefont
  {Rohlfing}}\ and\ \bibinfo {author} {\bibfnamefont {S.~G.}\ \bibnamefont
  {Louie}},\ }\href@noop {} {\bibfield  {journal} {\bibinfo  {journal} {Phys.
  Rev. B}\ }\textbf {\bibinfo {volume} {62}},\ \bibinfo {pages} {4927}
  (\bibinfo {year} {2000})}\BibitemShut {NoStop}%
\bibitem [{\citenamefont {Sethi}\ \emph {et~al.}(2021)\citenamefont {Sethi},
  \citenamefont {Zhou}, \citenamefont {Zhu}, \citenamefont {Yang},\ and\
  \citenamefont {Liu}}]{sethi2021flat}%
  \BibitemOpen
  \bibfield  {author} {\bibinfo {author} {\bibfnamefont {G.}~\bibnamefont
  {Sethi}}, \bibinfo {author} {\bibfnamefont {Y.}~\bibnamefont {Zhou}},
  \bibinfo {author} {\bibfnamefont {L.}~\bibnamefont {Zhu}}, \bibinfo {author}
  {\bibfnamefont {L.}~\bibnamefont {Yang}},\ and\ \bibinfo {author}
  {\bibfnamefont {F.}~\bibnamefont {Liu}},\ }\href@noop {} {\bibfield
  {journal} {\bibinfo  {journal} {Phys. Rev. Lett.}\ }\textbf {\bibinfo
  {volume} {126}},\ \bibinfo {pages} {196403} (\bibinfo {year}
  {2021})}\BibitemShut {NoStop}%
\bibitem [{\citenamefont {Xiao}\ \emph
  {et~al.}(2012{\natexlab{b}})\citenamefont {Xiao}, \citenamefont {Liu},
  \citenamefont {Feng}, \citenamefont {Xu},\ and\ \citenamefont
  {Yao}}]{xiao2012coupled}%
  \BibitemOpen
  \bibfield  {author} {\bibinfo {author} {\bibfnamefont {D.}~\bibnamefont
  {Xiao}}, \bibinfo {author} {\bibfnamefont {G.-B.}\ \bibnamefont {Liu}},
  \bibinfo {author} {\bibfnamefont {W.}~\bibnamefont {Feng}}, \bibinfo {author}
  {\bibfnamefont {X.}~\bibnamefont {Xu}},\ and\ \bibinfo {author}
  {\bibfnamefont {W.}~\bibnamefont {Yao}},\ }\href@noop {} {\bibfield
  {journal} {\bibinfo  {journal} {Phys. Rev. Lett.}\ }\textbf {\bibinfo
  {volume} {108}},\ \bibinfo {pages} {196802} (\bibinfo {year}
  {2012}{\natexlab{b}})}\BibitemShut {NoStop}%
\bibitem [{\citenamefont {Mahon}\ \emph {et~al.}(2019)\citenamefont {Mahon},
  \citenamefont {Muniz},\ and\ \citenamefont {Sipe}}]{mahon2019quantum}%
  \BibitemOpen
  \bibfield  {author} {\bibinfo {author} {\bibfnamefont {P.~T.}\ \bibnamefont
  {Mahon}}, \bibinfo {author} {\bibfnamefont {R.~A.}\ \bibnamefont {Muniz}},\
  and\ \bibinfo {author} {\bibfnamefont {J.~E.}~\bibnamefont {Sipe}},\ }\href@noop
  {} {\bibfield  {journal} {\bibinfo  {journal} {Phys. Rev. B}\ }\textbf
  {\bibinfo {volume} {100}},\ \bibinfo {pages} {075203} (\bibinfo {year}
  {2019})}\BibitemShut {NoStop}%
\bibitem [{\citenamefont {Sun}\ \emph {et~al.}(2006)\citenamefont {Sun},
  \citenamefont {Giebink}, \citenamefont {Kanno}, \citenamefont {Ma},
  \citenamefont {Thompson},\ and\ \citenamefont
  {Forrest}}]{triplet-stability-1}%
  \BibitemOpen
  \bibfield  {author} {\bibinfo {author} {\bibfnamefont {Y.}~\bibnamefont
  {Sun}}, \bibinfo {author} {\bibfnamefont {N.~C.}\ \bibnamefont {Giebink}},
  \bibinfo {author} {\bibfnamefont {H.}~\bibnamefont {Kanno}}, \bibinfo
  {author} {\bibfnamefont {B.}~\bibnamefont {Ma}}, \bibinfo {author}
  {\bibfnamefont {M.~E.}\ \bibnamefont {Thompson}},\ and\ \bibinfo {author}
  {\bibfnamefont {S.~R.}\ \bibnamefont {Forrest}},\ }\href@noop {} {\bibfield
  {journal} {\bibinfo  {journal} {Nature}\ }\textbf {\bibinfo {volume} {440}},\
  \bibinfo {pages} {908} (\bibinfo {year} {2006})}\BibitemShut {NoStop}%
\bibitem [{\citenamefont {Shuai}\ \emph {et~al.}(2000)\citenamefont {Shuai},
  \citenamefont {Beljonne}, \citenamefont {Silbey},\ and\ \citenamefont
  {Br{\'e}das}}]{triplet-stability-2}%
  \BibitemOpen
  \bibfield  {author} {\bibinfo {author} {\bibfnamefont {Z.}~\bibnamefont
  {Shuai}}, \bibinfo {author} {\bibfnamefont {D.}~\bibnamefont {Beljonne}},
  \bibinfo {author} {\bibfnamefont {R.~J.}~\bibnamefont {Silbey}},\ and\ \bibinfo
  {author} {\bibfnamefont {J.~L.}~\bibnamefont {Br{\'e}das}},\ }\href@noop {}
  {\bibfield  {journal} {\bibinfo  {journal} {Phys. Rev. Lett.}\ }\textbf
  {\bibinfo {volume} {84}},\ \bibinfo {pages} {131} (\bibinfo {year}
  {2000})}\BibitemShut {NoStop}%
\end{thebibliography}

\section{Supplemental Material (SM)}

\begin{figure}[htb]
\includegraphics[scale=0.35]{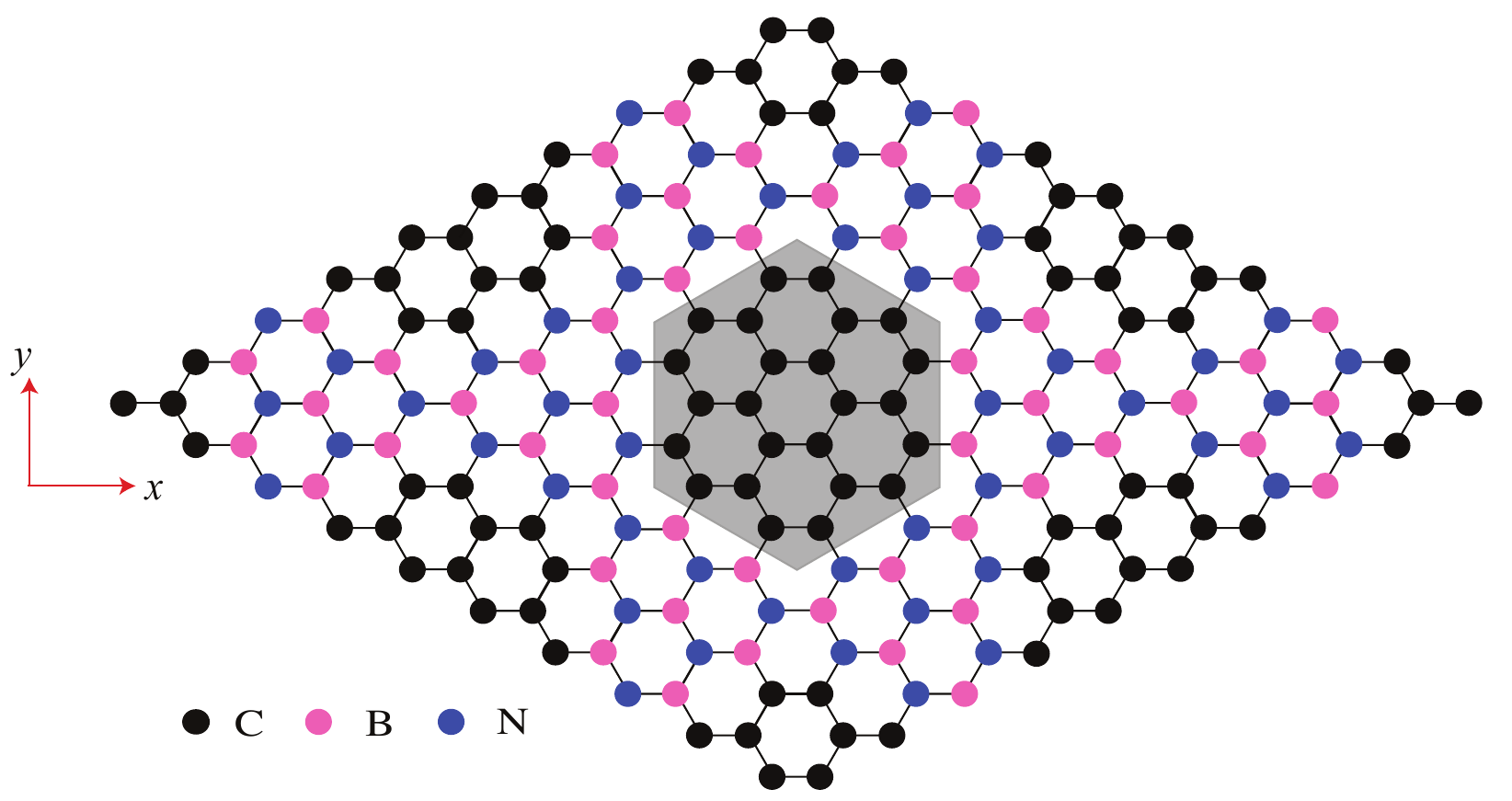}
\caption*{\textbf{Figure S1} : The lattice structure of hBNK-G in 2x2 supercell representation. The void areas of the rhombus unit cell of hBNK (see Fig.1a in the main text) are filled with hexagonal graphene domain (shaded hexagon) to form the lattice structure of hBNK-G.}
\end{figure}

\subsection{Density Functional Theory (DFT) Calculation}
Our DFT calculations are performed with the projector-augmented wave pseudopotentials and the generalized gradient approximation (GGA) within Perdew-Burke-Ernzerhof (PBE) exchange and correlation functional using Vienna Ab initio Simulation Package (VASP) code
[1-3]. An energy cut-off of 400 eV and a 12 x 12 x 1 Monkhorst-pack 
\textbf{k}-point grid is used for the relaxation. The structure is optimized until the atomic-forces are smaller than 0.01 eV/\AA. 
The vacuum layer is set to 20 \AA~ in non-periodic direction to avoid any interaction between adjacent unit cells. We consider BZ sampling over 24 x 24 x 1 to study the PBE based electronic properties. For the Berry curvature related calculations, we use the Wannier90 package [4].
\subsection{Charge Carrier Mobility Calculations}
We calculate the acoustic phonon mediated charge carrier mobility of a 2D material using deformation potential (DP) theory [5]. According to the DP theory, expression of charge carrier mobility at any temperature T is given by
\begin{equation}
\mu_{2D} = \frac{2e\hbar^3C_{2D}}{3k_BT|m^*|^2E^2_I}    
\end{equation}
where $\hbar$ is the reduced Plank constant, $K_B$ is the Boltzmann constant, $e$ is the electronic charge, and the temperature $T$ is set as room temperature. The $C_{2D}$ is the in-plane stiffness under the applied strain ($\delta$, in percentage) along a particular direction, \textit{i.e.}, $x$ or $y$. Here we apply the strain only along the $x$-direction since the hBNK-G structure is isotropic in nature. It takes the following form:
\begin{equation}
C_{2D} = \frac{1}{A_0}\frac{\partial^2E_{Total}}{\partial\delta^2}
\end{equation}
where $E_{Total}$ is the total energy of the unit cell and $A_0$ is the area of the unit cell within equilibrium conditions. The deformation potential, $E_I$ is the change of conduction band minima (CBM) position with applied strain. It can be expressed as: 
\begin{equation}
    E_I = \frac{\partial E_{CBM}}{\partial\delta}
\end{equation}
The effective mass of the electron with zero-strain condition is given by 
\begin{equation}
    m^* = \frac{\hbar^2}{\partial^2\epsilon(k)/\partial k^2}
\end{equation}
The calculated values of these parameters are listed in Table II. The variation of these quantities with the applied strain are plotted in Fig.S2.
\begin{table}[h]
\caption{\label{tab:table2}
Area of the unit cell ($A_0$) under equilibrium conditions, effective mass of electron ($m^*$) in unit of free electron mass, in-plane stiffness ($C_{2D}$) and deformation potential ($E_I$) of hBNK-G.}
\begin{tabular}{|c|c|c|c|c|}
\hline
System name &$A_0$(SI unit) &$m^*$ &$C_{2D}$ (SI unit)& $E_I$ (SI unit)\\
\hline
hBNK-G &135x10$^{-20}$  & 0.419 &354.4 & 7.224x10$^{-20}$\\ 
\hline
\end{tabular}
\end{table}

\begin{figure}[htb]
\includegraphics[scale=0.45]{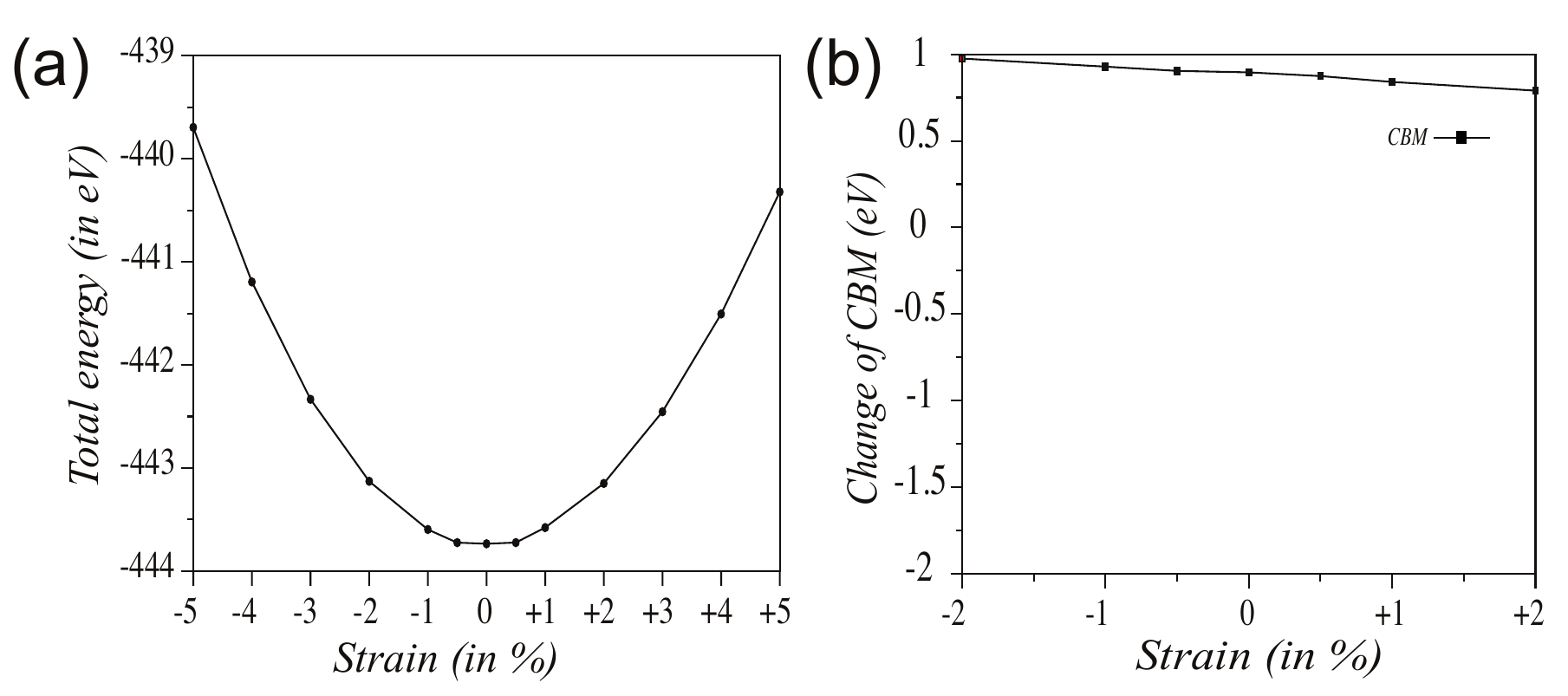}
\caption*{\textbf{Figure S2} : (a) Variation of total energy as a function of in-plane strain. (b) Change of CBM with applied strain of hBNK-G. Fermi energy is set to zero for band edge plots.}
\end{figure}

\clearpage

\subsection{Berry curvature of hBNK-G in irreducible BZ}

\begin{figure}[htb]
\includegraphics[scale=0.5]{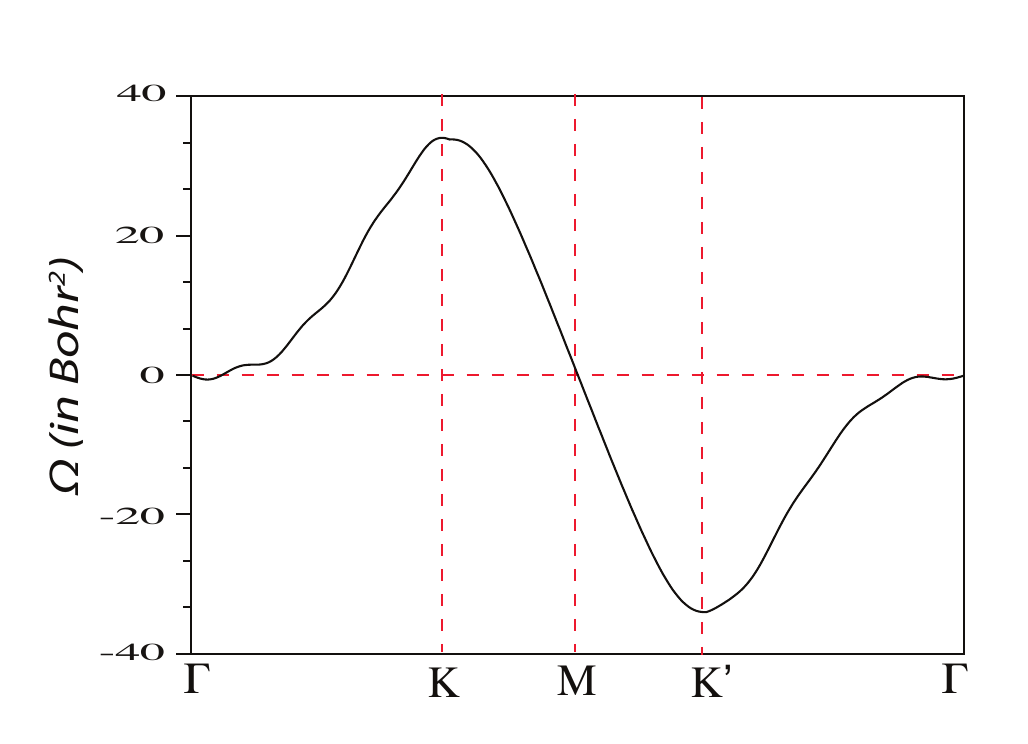}
\caption*{\textbf{Figure S3} : Berry curvature of hBNK-G in the irreducible BZ. Highest value of Berry curvature is $|36|$ Bohr$^2$.}
\end{figure}

\subsection{Low-lying singlet excitonic eigenvalues of hBNK-G}

\begin{figure}[htb]
\includegraphics[scale=0.5]{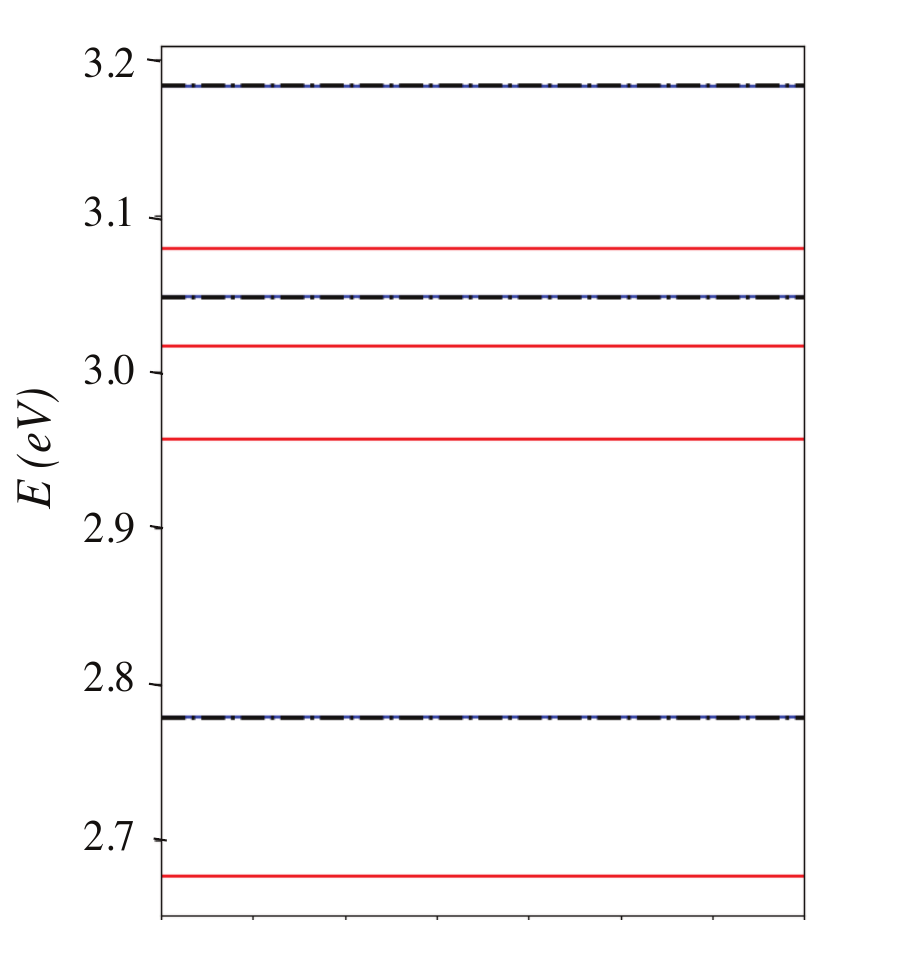}
\caption*{\textbf{Figure S4} : The singlet-excitonic energy levels of hBNK-G, as obtained by solving BSE. Blue (red) lines represent bright (dark) excitons, that are determined based on their relatively large (small) oscillator strengths. Here, the excitons with less than 5\% of the oscillator strength of the brightest exciton are defined as dark excitons. The black dashed lines represent degenerate excitonic states of either the bright or dark excitons.}
\end{figure}

\subsection{Convergence test of $GW$-BSE calculations}
The first-principle based $GW$ plus BSE calculations are performed using the BerkeleyGW package [6-8]. The DFT calculations are performed using the Quantum Espresso code for the $GW$ and BSE calculations [9]. For the $GW$ calculations, the plane-wave energy cutoff is set at 50 Ry based on norm-conserving pseudopotentials that are taken from the Pseudo-Dojo database [10]. We consider a 8 x 8 x1 coarse \textbf{k}-grid for the $GW$ calculations. A slab truncated Coulomb potential method is used to accelerate the convergence [11]. The quasiparticle band gap is calculated within the plasmon pole model [7]. We use 32 x 32 x 1 fine \textbf{k}-grid for the optical absorption with 3 valence bands and 4 conduction bands. The polarization of the incident light is considered to be parallel to the 2D plane of hBNK-G. We consider zero center-of-mass momentum of excitons in our BSE computation.\\

\textbf{$GW$ convergences:}\\

We first achieve the convergence by varying the total number of bands, comprising of a few valence bands and a few conductions bands to calculate the dielectric matrix with a k-point mesh of 5 x 5 x 1, dielectric matrix cut-off of 8 Ry, and 50 Ry bare Coulomb cut-off, 
as shown in Table III. Then we fix the number of bands at 615 and perform the convergence test for the dielectric matrix cut-off with 5 x 5 x 1 coarse \textbf{k}-point as shown in Table III. With a fixed
dielectric matrix cut-off at 8 Ry, we further test the convergence for the bare Coulomb cut-off with same number of bands and coarse \textbf{k}-point grid, as shown in Table III. We find 50 Ry as the 
converged bare Coulomb cut-off. Finally, we test for the coarse \textbf{k}-point grid for the $GW$ calculations, as shown in Table III. We find 8 x 8 x 1 coarse \textbf{k}-point grid as a converged coarse \textbf{k}-grid. For this entire convergence test we set 1 meV tolerance for the $GW$ gap.

\begin{table}[h]
\caption*{\textbf{Table S1} :
N$_b$ is the number of bands (occupied plus unoccupied), E$_g$ in (eV) is the $GW$ gap, Encut (in Ry) is the energy cut-off of the plane wave, Epsilon-cut (in Ry) is the dielectric matrix cut-off in 
the epsilon calculations.}
\begin{tabular}{|c|c|c|c|c|c|c|c|}
\hline
N$_b$ & E$_g$ & Epsilon-cut & E$_g$ & Encut & E$_g$ & \textbf{k}-point & E$_g$\\
\hline
504 &3.36 & 8& 3.36 & 30 & 3.35 & 5x5x1 & 3.36\\ 
\hline
615 &3.36 & 9& 3.36 & 40 & 3.36 & 8x8x1 & 3.50\\ 
\hline
712 &3.36 & 10& 3.36 & 50 & 3.36 & 10x10x1 & 3.50\\ 
\hline
\end{tabular}
\end{table}

\clearpage

\textbf{BSE convergences:}
We test the convergence of BSE calculations on the number of valence bands and conduction bands in the optical calculations. We find 3 valence bands and 4 conduction bands are the converged parameters in the BSE calculation of hBNK-G. Then we do test for the fine \textbf{k}-grid in the BSE calculations. We find 32 x 32 x 1 fine \textbf{k}-grid is the converged parameter. We plot the absorption spectra for the singlet exciton of hBNK-G in the Fig.S5.

\begin{figure}[htb]
\includegraphics[scale=0.35]{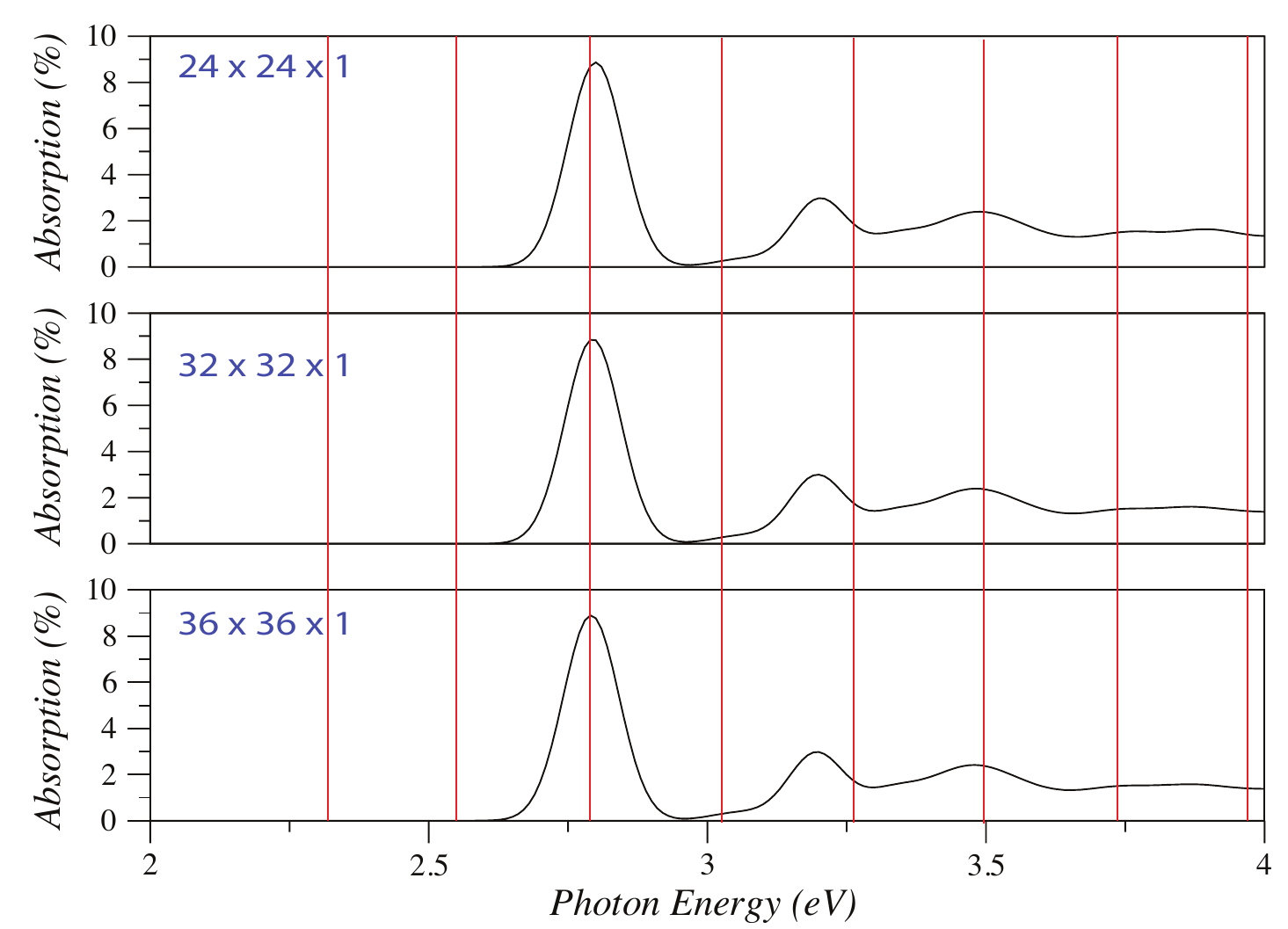}
\caption*{\textbf{Figure S5} : BSE convergence test on fine \textbf{k}-kgrid with 3 valence bands and 4 conduction bands. The 
top, middle and bottom panels represent the absorption spectra of hBNK-G, as obtained from \textbf{k}-grids of 24 x 24 x 1, 32 x 32 x 1 and 36 x 36 x 1, respectively.}
\end{figure}

\clearpage

\subsection{Original plot of Fig.3 in the main text}
Here we provide the original plots of Fig.3 in the main text, which we obtain using a python code that is given in the github repository (bgwtools) of BerkeleyGW package [5]. The link of the same is \url{https://github.com/BerkeleyGW}. For better representation, we rotate all the figures by 45-degree angle, so that the first Brillouin zone appears as a perfect hexagon, as it should be and present the same in Fig.3 in the main text.

\begin{figure}[htb]
\includegraphics[scale=0.3]{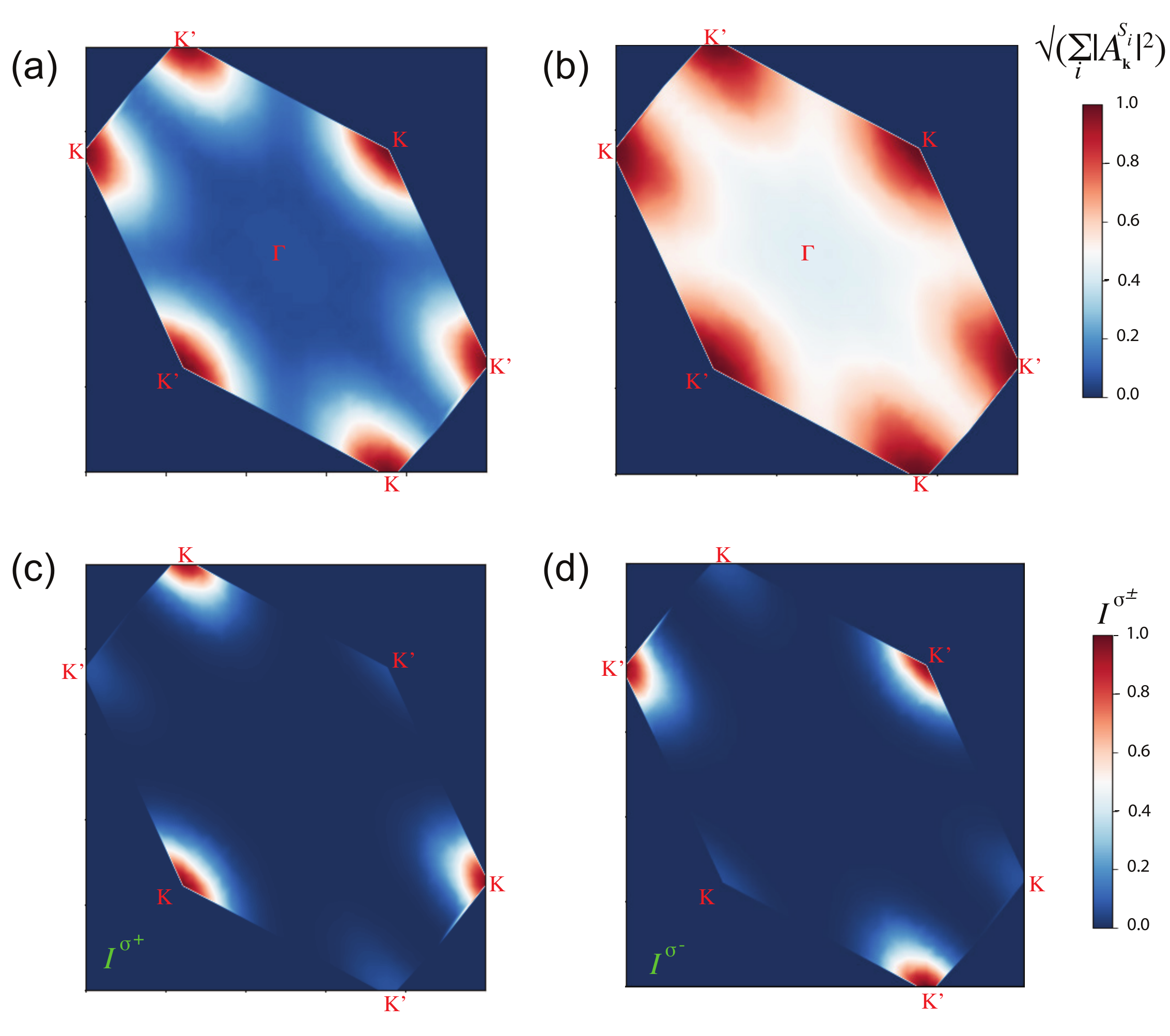}
\caption*{\textbf{Figure S6} : The square root of the sum of the squared excitonic envelope function (defined in the 
main text) of (a) the first bright singlet ($A$) and (b) the first triplet ($B$) exciton of hBNK-G over the hexagonal first BZ. The color bar displays the normalized value of the same. The oscillator 
strengths of the first bright singlet ($A$) exciton in presence of (c) left- and (d) right-handed circular polarized lights. The color bar depicts the normalized values of the oscillator strength.}
\end{figure}

\clearpage

\subsection{Envelope function of the first triplet exciton}

\begin{figure}[htb]
\includegraphics[scale=0.35]{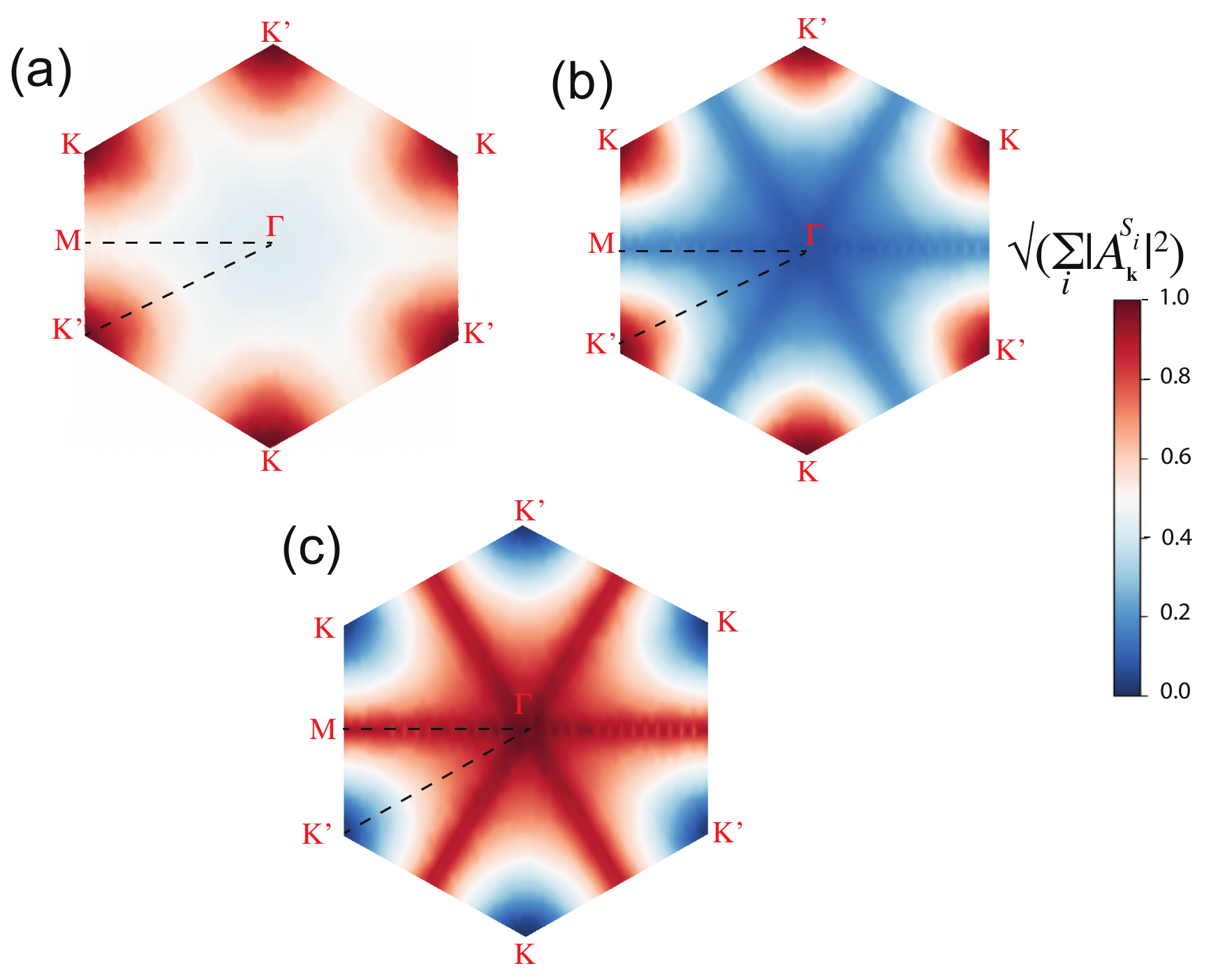}
\caption*{\textbf{Figure S7} : The square root of the sum of the squared excitonic envelope function of first triplet 
exciton of hBNK-G with contributions from (a) converged number of bands with 3 valence bands and 4 conduction bands and from (b) the band edges, \textit{i.e.}, 1 valence band (top most) and 1 conduction band (bottom most). (c) The resultant values after subtracting the values of plot (b) from plot (a). Therefore, it represents the contribution from other 2 valence and other 3 conduction bands other than the band edges.}
\end{figure}

\subsection{ References in SM}
[1] G. Kresse, and J. Furthmuller, Efficiency of ab-initio total energy calculations for metals and semiconductors using a plane-wave basis set, Compu. Mater. Sci. \textbf{6}, 15-50 (1996).\\

[2] G. Kresse and J. Hafner, Ab initio molecular dynamics for
open-shell transition metals, Phys. Rev. B \textbf{48}, 13115 (1993).\\

[3] J. P. Perdew, K. Burke, and M. Ernzerhof, Generalized gradient approximation made simple, Phys. Rev. Lett. \textbf{77}, 3865 (1996).\\

[4] G. Pizzi et. al., Wannier90 as a community code: new features and applications, J. Phys.: Conden. Matter \textbf{32}, 165902 (2020).\\

[5] J. Bardeen and W. Shockley, Deformation potentials and mobilities in non-polar crystals, Phys. Rev. \textbf{80}, 72 (1950).\\

[6] J. Deslippe, G. Samsonidze, D. A. Strubbe, M. Jain, M. L. Cohen, and S. G. Louie, Berkeleygw a massively parallel computer package for the calculation of the quasiparticle and optical properties of materials and nanostructures, Computer Physics Communications \textbf{183}, 1269 (2012).\\

[7] M. S. Hybertsen and S. G. Louie, Electron correlation in semiconductors and insulators: Band gaps and quasiparticle energies, Phys. Rev. B \textbf{34}, 5390 (1986).\\

[8] M. Rohlfing and S. G. Louie, Electron-hole excitations and optical spectra from first principles, Phys. Rev. B \textbf{62}, 4927 (2000).\\

[9] P. Giannozzi, S. Baroni, N. Bonini, M. Calandra, R. Car, C. Cavazzoni, D. Ceresoli, G. L. Chiarotti, M. Cococcioni, I. Dabo, et al., Quantum espresso a modular and open-source software project for quantum simulations of materials, Journal of physics: Condensed matter \textbf{21}, 395502 (2009).\\

[10] M. J. van Setten, et. al., The PseudoDojo: Traning and grading a 85 element optimized normconversing pseudopotential table, Computer Physics Communications \textbf{226}, 39-54 (2018).\\

[11] S. Ismali-Beigi, Truncation of periodic image interactions for confined systems, Phys. Rev. B \textbf{73}, 233103 (2006).

\end{document}